\documentclass{ametsocV6.1}
\usepackage{float}
\usepackage{rotating}

\title{Formation and evolution of turbulence \\ in convectively unstable internal solitary waves of depression \\ shoaling over gentle slopes in the South China Sea}

\authors{Tilemachos Bolioudakis,\aff{a}\correspondingauthor{Tilemachos Bolioudakis, tb424@cornell.edu} 
Theodoros Diamantopoulos,\aff{a} 
Peter J. Diamessis,\aff{a}  
Ren-Chieh Lien,\aff{b} 
Kevin G. Lamb,\aff{c}
Gustavo Rivera-Rosario,\aff{a} 
Greg N. Thomsen,\aff{d}
}

\affiliation{\aff{a}{School of Civil and Environmental Engineering, Cornell University, Ithaca, New York}\\
\aff{b}{Applied Physics Laboratory, University of Washington, Seattle, Washington}\\
\aff{c}{Department of Applied Mathematics, University of Waterloo, Waterloo, Canada}\\
\aff{d}{Wandering Wakhs Research, Austin, Texas}
}

\abstract{The shoaling of high-amplitude Internal Solitary Waves (ISWs) of depression in the South China Sea (SCS) is examined through large-scale parallel turbulence-resolving high-accuracy/resolution simulations. A select, near-isobath-normal, bathymetric transect of the gentle SCS continental slope is employed together with stratification and current profiles obtained by in-situ measurements. Three simulations of separate ISWs with initial deep-water amplitudes in the range [136m, 150m] leverage a novel wave-tracking capability for a propagation distance of 80km and accurately reproduce key features of in-situ-observed phenomena with significantly higher spatiotemporal resolution. The interplay between convective and shear instability and the associated turbulence formation and evolution, as a function of deep-water ISW amplitude are further studied in-part revealing processes previously not observed in the field. Across all three waves, the convective instability develops in a similar fashion. Heavier water entrained from the wave rear plunges into its interior, giving rise to transient, yet distinct, subsurface vortical structures. Ultimately, a gravity current is triggered which horizontally advances through the wave interior and mixes it down to pycnocline’s base. Although the waveform remains distinctly symmetric, Kelvin-Helmholtz billows emerge near the well-mixed ISW trough, disturb the wave’s trailing edge and give rise to an active wake. The evolution of the kinetic energy associated with finer-scale perturbations to the ISW-induced velocity field shows two different growth regimes, each dominated by either convective or shear instability. The wake's perturbation kinetic energy is nonlinearly dependent on deep-water wave amplitude and can become a sizable fraction of the kinetic energy of the deep-water ISW.}

\begin{document}

\maketitle

%
\section{Introduction}
\label{sec:intro}

Internal solitary waves (ISWs) are ubiquitous oceanic phenomena found on continental slopes and shelves, in submarine canyons, and over oceanic topographic features (e.g. \citealt{sandstrom1984,klymak2003,scotti2004,carter2005,moum2003,moum2007a,shroyer2011,lien2005,lien2012,lien2014,zhang2015,cheng2024}). Oceanic ISWs typically result from the nonlinear dispersive evolution of the internal tides (\citealt{helfrich2006,Stastna2022}). Most commonly occurring as waves of depression, ISWs are nonlinear, and non-hydrostatic waves, characterized by a large vertical displacement of the pycnocline. Due to the balance between nonlinear steepening and physical dispersion, ISWs maintain their waveform while propagating over long distances \citep{jackson2012}. One of their defining properties is their ability to efficiently transport mass and energy. Moreover, such waves have also been observed to become unstable as they propagate over varying bathymetry, resulting in mixing and energy dissipation far from their generation site \citep{lamb2019}. Additionally, ISWs are efficient suppliers of nutrients into the upper ocean, as they transport the particulate matter they entrain over long distances towards the continental shelf and exchange it with the surrounding water (\citealt{lamb2002,lien2012}). Consequently, they affect the heat, salt, nutrient, and mass fluxes between the open ocean and coastal waters, and are therefore crucial for primary production and marine ecology (\citealt{sandstrom1984,moore2007}).

\par 
This study is motivated by field observations in the South China Sea \citep{lien2012,lien2014} of convectively unstable mode-1 shoaling non-linear internal waves in which a subsurface recirculating core was distinctly observed to form. The observations of \cite{chang2021b} confirmed the hypothesis of \cite{lien2012,lien2014} that convective instability of ISWs is a systematically occurring phenomenon in this particular region. Convective instability is generated when the maximum wave-induced velocity exceeds the wave celerity \citep{lamb2002}. The subsurface recirculating core can be described as a region with closed streamlines \citep{aigner1999}, computed in a frame of reference moving with the wave (\citealt{lamb2003};\citealt{riverarosario2020}), where the maximum wave-induced horizontal velocity occurs well below the surface. For a subsurface core to form, the sign of the pre-existing vorticity of the near-surface background current has to be opposite of that generated by the propagating wave \citep{he2019}. In addition, the size of the convectively unstable region generated is largely determined by the magnitude of the near-surface transverse vorticity, while the interaction with the gently varying bathymetric slope accelerates the subsurface core formation \citep{riverarosario2020,riverarosario2022}. Numerical studies have shown that, during the onset of convective instability, heavier fluid from the rear of the wave plunges forward on top of lighter fluid as it enters the wave interior, producing an overturning and recirculating pattern. However, this subsurface instability, which is confined to the wave interior, is not abrupt enough to cause a complete wave disintegration, as observed during ISW breaking where the pycnocline intersects the slope (\citealt{helfrich1992};\citealt{vlasenko2005b}). Instead, the ISW maintains a nearly symmetric waveform as it continues to shoal towards the continental shelf \citep{lien2014}.

\par
The convectively unstable core is linked to enhanced turbulent mixing and energy dissipation inside the ISW \citep{lien2012}. In the meantime, the core has also been found to exchange its water mass with its environment via entrainment into the recirculating core and detrainment through a "leaky" behavior strictly linked with the tail of the wave, with what was inferred to be a trailing turbulent wake (\citealt{moum2003,lien2012}).

\par
Shear instability and Kelvin–Helmholtz (K–H) billows were first observed with biosonic measurements within the interior of a mode-1 ISW on the Oregon shelf by \cite{moum2003}. The authors report shear instability initiated both at the leading edge and the trough of observed ISWs. Recent observations have suggested that shear instability can also originate in convectively-unstable waves through the resulting mixing of the wave core and weakening the stratification therein, namely at the wave trough \citep{chang2021b}, implying the dynamice coupling between convective and shear instabilities within ISWs.
Moreover, the accompanying enhancement of the vertical shear is generally found to be stronger along the lower periphery of the wave core, while stronger stratification is located predominantly between the periphery of the wave core and the pycnocline \citep{carr2011}, a combination that works favorably for shear instability to occur \citep{chang2021b}. In contrast to \cite{moum2003}, both \cite{chang2021b,chang2021c} and \cite{lamb2019} report K-H billows to mainly emerge near the trough of the wave, where the Richardson number ($Ri=N^2/S^2$ ; $N$ and $S$ are the Brunt-Väisälä frequency and the vertical shear)  is less than a value 1/4. Note that controlled laboratory studies \citep{carr2008b} and direct numerical simulations \citep{barad2010,carr2011} indicate a more stringent critical value of 0.1. The billows grow while traveling toward the ISW rear along the wave interface until their collapse, a process that can non-trivially modify the observed symmetric waveform.

\par
Clear signatures of shear and convective instabilities within ISWs \citep{lamb2019} have not been yet observed in-situ in the ocean due to the  intermittency and rapid evolution of such phenomena in conjunction with the intrinsic limitations in instrumentation and measurement sampling/resolution \citep{chang2021b}. As such,  a detailed representation of the two-dimensional structure of these instabilities and their subsequent evolution on the along-wave/depth plane at sufficient spatial and temporal resolution has not yet been offered. 

\par
Caution is, however, warranted when attempting to leverage the, otherwise illuminating, insights of various laboratory (e.g. \citealt{troy2005,carr2008b,fructus2009}) and numerical studies (e.g. \citealt{barad2010,carr2011,lamb2011,carr2012,stastna2024}) towards interpreting the physics of the large-amplitude ISWs observed in the SCS. The SCS (and other in-situ) waves reside in a different regime of the parameter space than their above laboratory/numerical counterparts over variable bathymetry which focus on internal wave instability over {\em {steep}} slopes ($>5\%$) and are primarily limited to the laboratory-scale. Some of the above studies consider uniform-depth water, accounted for in numerical simulations by considering a wave-fixed frame of reference to minimize computational cost, which obviously precludes the consideration of variable bathymetry and its impact on the ISW waveform \citep{lamb2019}. 

\par
The tracking of ISW propagation over {\em {gentle}} slopes (such as those of the SCS) while capturing the associated instability-driven localized turbulence, both of which are of interest to this study,  has not yet been possible in the laboratory or through numerical simulations due to the lack of sufficiently long facilities or the infeasibility of the large computational grid required. Moreover, the range of turbulent scales attainable in the laboratory or through numerical modeling is significantly restricted with respect to the ocean \citep{lamb2019}. 

\par
Focusing now on ocean process modeling, the simulation of the shoaling of ISWs in the SCS while resolving the resulting instabilities and (at least the larger scales of the) highly energetic turbulence in the wave interior, as reported by \cite{lien2012,lien2014,chang2021b}, poses a state-of-the-art challenge.  To this end, the characteristic lengthscales of interest extend from ($\mathcal{O}(100 \:$ km) propagation distances and transition down to the wavelength ($\mathcal{O}(1 \:$ km), as calculated according to \citet{koop1981} and \citet{riverarosario2022}) and the scales of the convectively-driven overturn ($\mathcal{O}(100 \: $m)), K-H billows ($\mathcal{O}(10 \: $m)) and finally turbulence ($\mathcal{O}(1 \: $m)).

\par 
The spectral-element method (SEM) is optimally suited for the above multiscale problem, as it can adequately capture the wave-scale response to the bathymetry and resulting finer-scale phenomena without the spurious influence of numerical dispersion and dissipation \citep{diamantopoulos2022}. One can also leverage the SEM's ability to localize resolution within the wave core in the vertical and progressively reduce it in the along-wave direction as the ISW approaches the convective instability location. Moreover,  central to enabling the use of the above advantages of the SEM to the particular flow phenomena of interest is the recent development of high-order-accurate-element-focused computationally efficient non-hydrostatic pressure solvers \citep{joshi2016a,diamantopoulos2022}. These numerical tools have finally  overcome long-standing challenges linked to high-aspect ratio grids and pressure-driven coupling over long domains \citep{Scotti:08}, which had persistently hampered the efficient performance of these solvers in the past.

\par 
This study uses the SEM-based flow-solver of \cite{diamantopoulos2022} to build on the previous work of \citet{riverarosario2020,diamantopoulos2021} and \citet{riverarosario2022} and reports on novel high-accuracy-resolution and fully nonlinear/non-hydrostatic turbulence-resolving simulations of shoaling ISWs over the SCS bathymetry in three dimensions. These simulations, which leverage non-trivial computational resources, are motivated and justified by the limited existing in-situ observations on convective and shear instabilities in ISWs shoaling over gentle slopes. The high-accuracy/resolution numerical modeling effectively enables the desired, and currently lacking, detail of representation on the structure and evolution of both of the above instabilities in shoaling ISWs while revealing phenomena previously not captured in the field.

\par
The primary objective of this effort is to computationally investigate the response of three high-amplitude ISWs (amplitude is defined as the deep-water maximum isopycnal displacement), at scales of the wave and finer, when shoaling over realistic gentle slopes in the SCS with in-situ-sampled background current/stratification profiles. In particular, this study aims to examine the two sequentially occurring instabilities in shoaling ISWs, focusing first on convective instability and, through the resulting weakening of stratification inside the ISWs, the transition to shear instability. By leveraging high spatiotemporal resolution and wave-tracking capabilities —further supported by supplementary animations— this work offers an unprecedented detail/resolution of the formation and evolution of these instabilities. An improved understanding of the structure and dynamics of the particular processes which were challenging, if not impossible, to be captured by field observations is thus enabled.

\par Three-dimensional visualizations enhance one's perspective on the structure of the processes of interest and their lateral variability, while a novel feature of the convective instability is further explored: a gravity current-like structure which horizontally propagates through the wave interior. This flow feature is quantitatively assessed in terms of the propagation of its front, and contrasted to its idealized lock-exchange counterpart at laboratory-scale. Linkages in the analysis of the two instabilities, and their subsequent evolution, are further buttressed through exploring the  kinetic energy of the perturbations (PKE) to the instantaneous traverse-averaged velocity field, a surrogate for the kinetic energy of turbulence. The spatial distribution of PKE within the ISW at select locations along the wave propagation path is contrasted in terms of the two instabilities, as is also the temporal evolution of the wave-integrated PKE along the propagation track. The study culminates with a novel presentation of the Kelvin-Helmholtz-billow-driven wake, in terms of coherent flow structures and along-wake distribution of the PKE. In regards to the latter topic, levels of shear-instability-driven PKE can visibly surpass those driven by convective instability, even at the most energetic stages of evolution of the latter. This novel finding points to the potential role of shear-instability-driven mixing in driving a longer-term (after the passage of the ISW) modification of the background water column.

\par
This paper is structured as follows: Sec. 2 describes the oceanic region of interest and field conditions that dictate the configuration of the computational domain. The governing equations and the numerical method are also reviewed. Sec. 3,~4,~5 and 6 analyze and inter-compare the turbulence-resolving simulations using both qualitative and quantitative metrics. Finally, Sec. 7 contains the summary and discussion of the overall findings and open questions.

\section{Method}
\label{sec:Method}

\subsection{Problem geometry}
\label{subsec:Geometry}

\begin{figure}[h!]
 \centerline{\includegraphics[width=33pc]{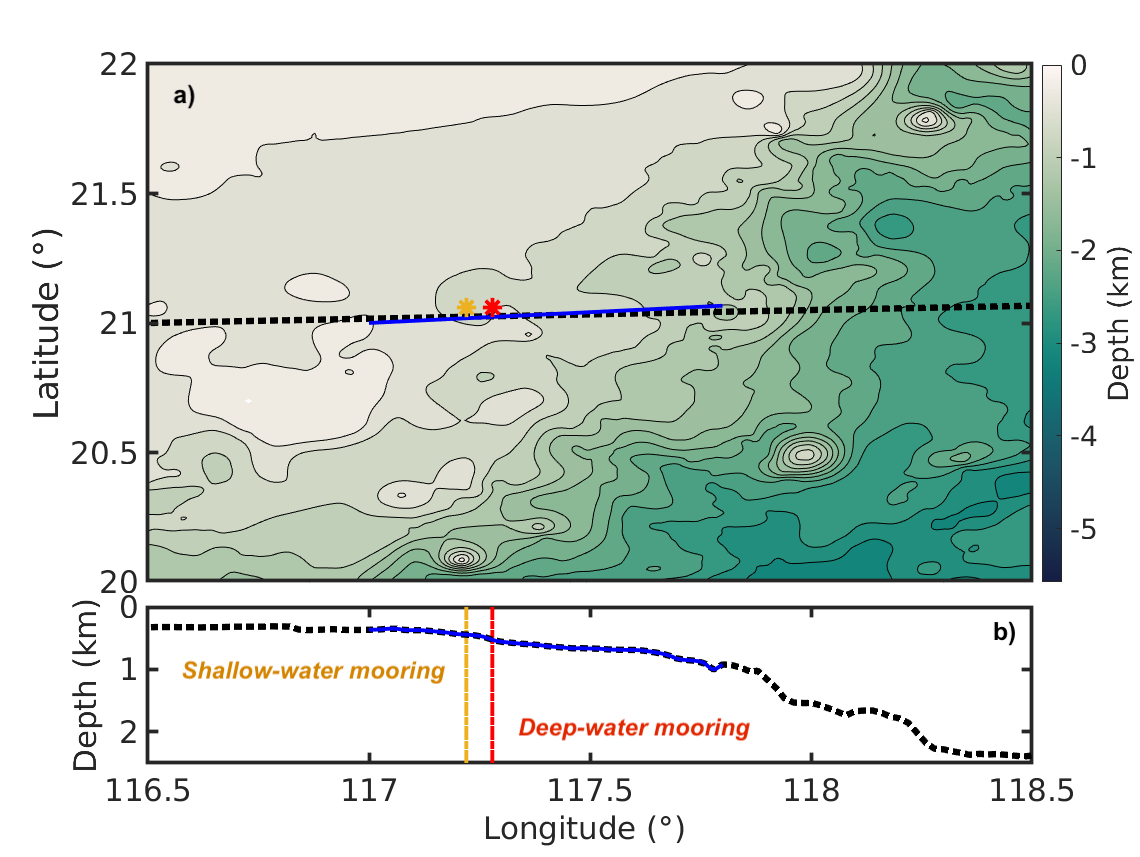}}
  \caption{Bathymetry map of the northern section of the South China Sea (a), in conjunction with the water depth along the transect (black dotted line in $b$). The bathymetric data for panel (a) are obtained from the General Bathymetric Chart of the Oceans (GEBCO). The aspect ratio of the map between longitude and latitude has been slightly adjusted from 1-to-1, to enhance visual clarity. The data actually used in the bathymetry of the high-resolution simulations reported here and shown in panel (b) are obtained from the high-accuracy shipboard echosounder measurements of \cite{lien2005}. The black dotted line on (a) the ship trajectory during the wave tracking in \cite{lien2014}. The solid blue line shows the region of this transect that utilized in the simulations reported here. Lastly, the {\em {shallow}} (orange star) and the {\em {deep}} moorings' (red star) locations are indicated, deployed 6km apart and covering the upper continental slope region. The colormap used is based on \cite{thyng2016}.}
  \label{fig:map}
\end{figure}

The broader area of interest is the South China Sea, which is a semi-enclosed marginal sea in the western Pacific Ocean characterized by basin depth initially greater than $3$km that gradually decreases from east to west, tracking the gentle slope up to the continental shelf. Figure \ref{fig:map} $(a)$, shows contours of the bathymetry of the region where the ISWs have been observed by \cite{lien2012,lien2014,chang2021a,chang2021b} and \cite{chang2021c}, along with the ship trajectory during their wave tracking in the experiment by \cite{lien2014} (black dotted line). The black contour lines (Fig.\ref{fig:map}a) highlight the principal bathymetric changes over the approximate transect ISWs were found to propagate. 

\par
The total length during the wave tracking measurements of \cite{lien2014} extended over a distance larger than 200 km. The high-accuracy/resolution numerical simulations reported here follow the approach of \cite{riverarosario2020,riverarosario2022}, and make use of bathymetric data sampled from the high-accuracy measurements of \cite{lien2005} (black dotted line of Fig. \ref{fig:map}$b$). In particular, the numerically investigated region focuses on the transect between 21$^{\circ}$N, 117.8$^{\circ}$E and 21$^{\circ}$N, 117.0$^{\circ}$E (80km in length), including the locations of both moorings {\em {shallow}} (21$^{\circ}$N, 117.22$^{\circ}$E) shown in orange color and {\em {deep}} (21$^{\circ}$N, 117.27$^{\circ}$E in red color) and indicated by the solid blue line in both Fig. \ref{fig:map} $(a)$ and $(b)$.
 
\par
The simulations use a Cartesian coordinate system obtained by converting latitude-longitude to UTM coordinates assuming negligible latitudinal changes along the ship track. At this point, one should note that in the three-dimensional computational coordinate system, $x$ is taken to represent the east-west direction, $z$ is the bottom-surface direction, and $y$ is the transverse (north-south) direction. In the simulation's framework, the $\hat{i}$ vector points towards the wave propagation direction (westward). Although the unitary vector $\hat{k}$ points from the bottom to the surface, all the figures presented in this paper use a positive sign on the vertical axis, leaving the "Depth" notation to denote distance below the surface.

\subsection{Field conditions} 
\label{subsec:Field_Cond}

\begin{figure}[h!]
 \centerline{\includegraphics[width=33pc]{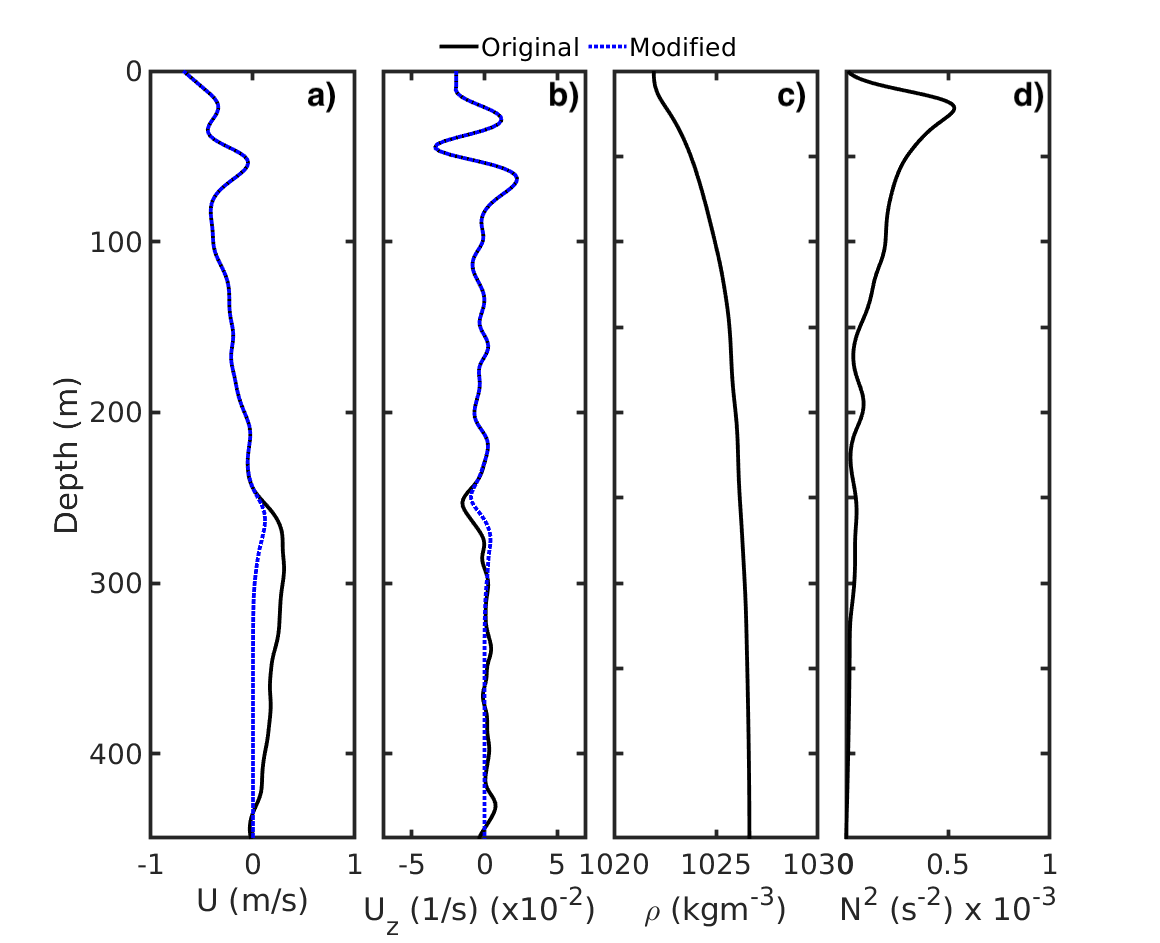}}
 \caption{Time-averaged vertical profiles of the background (a) current, (b) shear, (c) density, and (d) squared Brunt–Väisälä frequency used in this study. The velocity profile shown involves a change in sign, identically to \cite{riverarosario2020} with respect to that originally sampled and reported by \cite{lien2014}. The blue line shows the modified profiles used as background conditions in this numerical study, to ensure the absence of any potential hydraulic interaction of the background current with the gently varying bathymetry.}
 \label{fig:insitu_profs}
\end{figure}

In Fig. \ref{fig:insitu_profs}, the time-averaged background horizontal current velocity $(a)$, shear $(b)$, density $(c)$ and squared Brunt-Väisälä frequency $N^2$ $(d)$ profiles sampled, are shown with black lines. Note that, as in \cite{riverarosario2020}, the near-surface background velocity is set to be negative (of opposite sign to what is reported in \citealt{lien2014}) because in this study the positive direction is taken to be the westward (see also Sec \ref{sec:Method}\ref{subsec:Geometry}). Following \cite{riverarosario2020}, the background horizontal velocity profile in the top 10 meters of the water column is obtained from linear extrapolation upward, due to the absence of in situ data. The horizontal velocity is gradually set to zero below 300 m of depth, preventing any hydraulic effects associated with the interactions between the bottom fluid layer and the gentle slope. The resulting velocity and shear profiles used in this study are shown as the blue lines in Fig. \ref{fig:insitu_profs}(a) and (b). In addition, all four profiles are strictly one-dimensional, and thus no variation in either horizontal direction is considered. Moreover, assuming ISWs with propagation speeds comparable to the principal internal tide propagation speed, the background fields are held steady throughout the simulation. Last but not least, the depth where the maximum BV frequency is detected ($z_0 = -22$m), is regarded hereafter as the position of the pycnocline. The resulting value of density at this depth is set as the reference density ($\rho_0 \: = \: 1022.58 \: $kg$ \cdot $m$^{-3} $).

\subsection{Governing equations} 
\label{subsec:Gov_Eq}

In the present study, the mathematical model uses the incompressible Euler equations under the Boussinesq approximation (IEEB), 

\begin{equation}
\frac{\partial \textbf{u}}{ \partial t}  \: =  \:- \textbf{u} \cdot \nabla \textbf{u} \: - U \frac{\partial}{\partial x} \textbf{u} \: - w \frac{\partial U}{\partial z} \hat{i}  \: - \: \frac{1}{\rho_0} \nabla p'  - \: \frac{g}{\rho_0} \rho' \hat{k} \:  ~~,
\label{eq1}
\end{equation}
 
\begin{equation}
\frac{\partial \rho'}{ \partial t}  \: =  \:- \textbf{u} \cdot \nabla (\: \overline{\rho} (z) \:+\: \rho' \:) ~~,
\label{eq2}
\end{equation}

\begin{equation}
\nabla \cdot \textbf{u} \: =  \: 0 ~~,
\label{eq3}
\end{equation}

\noindent where, any possible effect of planetary rotation is neglected, as the shoaling effects are suggested to dominate over the depth of interest \citep{lamb2015}. Here $U(z)$ is a background current oriented along the wave propagation direction; it is assumed to be steady. One effectively solves for the three-dimensional vector $\textbf{u}$, in three-dimensional Cartesian coordinate space where $\textbf{x}$ is the vector of position therein, which consists of contributions from the two-dimensional wave-induced field and the subsequent turbulence.

\par
Accordingly, the density field is decomposed into a reference value $\rho_0$ (value taken from depth where $N^2$ is maximum), a steady background profile $\overline{\rho}(z)$ (obtained from the profiles measured in the water column) and a wave/turbulent perturbation  $\rho'$, such that $\rho \:=$  $\rho_0$ + $\overline{\rho}(z)$ + $\rho'(\textbf{x},t))$ with $\rho_0 \gg \overline{\rho}(z) \gg \rho'(\textbf{x},t))$. The hydrostatic balance between the reference pressure, away from the wave, and the background density has been subtracted from Eq. (\ref{eq1}); $p'$ represents the corresponding pressure perturbation.

\par
Neglecting the viscous and diffusive terms is based on the assumption that the smallest-resolved physical lengthscales are far larger than the Kolmogorov and Batchelor scales. In addition, any potential viscous effects at the bed due to a no-slip boundary condition and the associated turbulent boundary layer, are assumed to negligibly impact the large-scale wave structure and the convective instability and are outside of the scope of this study.
\par
The governing equations of the mathematical model are inviscid and non-diffusive. However, the numerical solution is subject to spectral filtering of both the velocity and density fields (see Appendix Aa). The spectral filtering is a computationally efficient surrogate of the insertion of a hyperviscous operator, typically used in purely Fourier solvers \citep{winters:04}, which can be flexibly applied to higher-order element-based techniques such as the one used here (see Sec. \ref{sec:Method}\ref{subsec:Num_Method}\ref{subsubsec:NSE_FGM} and \citealt{ozgokmen:09}). As elaborated in \citet{winters:15} (cf. their Section 2), when combined with a high-order-accurate method, the use of spectral filtering provides numerical stability against aliasing over a focused range of the smallest resolved scales in the flow while reproducing nearly inviscid and non-diffusive  physics of the larger flow scales, i.e., those on which the spectral filter does not act. Finally, since mixing is essentially effected at the smallest-resolved scales on the computational grid, and not at the actual Batchelor scale, we elect to avoid any associated quantitative discussions in this paper.
\par
With the exception of the periodic transverse direction, no boundary conditions are formally enforced in the velocity and density fields as these would be applied during the treatment of the viscous and diffusive terms \citep{diamantopoulos2022}, which are absent in the governing equations, Eqs. (\ref{eq1})-(\ref{eq3}). At the left and right boundaries of the overlapping window approach used to advance the ISW along the shoaling track (Sec. \ref{sec:Method}\ref{subsec:Sim_descr}\ref{subsubsec:Comp_domain}), which include the deep and shallow water boundaries of the computational domain, inflow/outflow is possible through the inserted the background current. No background current normal to the top/bottom boundaries exists, whereas instabilities and turbulence occur sufficiently offset from the top surface and bed, rendering these boundaries effectively impermeable. Constant inflow/outflow at the left/right boundaries and impermeability of the top and bottom ones are both consistently built into the nonhomogeneous Neumann boundary conditions  applied to the pressure on the $x-z$ plane \citep{karniadakis1991}, which implicitly assume a constant boundary-normal velocity \citep{lloret:24b}

\subsection{Numerical Method} 
\label{subsec:Num_Method}

\subsubsection{Generating the initial conditions} 
\label{subsubsec:Gov_Eq}

\par
The initial two-dimensional ISW is generated identically to \cite{riverarosario2020}. The fully nonlinear Dubreil-Jacotin-Long (DJL) equation (\citealt{long1953,turkington1991}) is solved using the pseudospectral method of \cite{dunphy2011} for the observed background conditions (Fig. \ref{fig:insitu_profs}) and prescribed available potential energy value. The resulting isopycnal displacement field $\eta(x,z)$ is used to construct the associated density $\rho'$ and velocity ($u \: \& \: w$) fields. The wave-representing fields are then interpolated onto the prescribed computational grid and finally inserted over a deep-water artificial constant-depth region (Fig. \ref{fig:domain}a) to prevent any potential spurious shoaling effects during the initialization.

\subsubsection{Hybrid Nodal Spectral Element / Fourier numerical method} 
\label{subsubsec:NSE_FGM}

\par
The numerical tool used in this study is a hybrid Nodal-Spectral-Element/Fourier-Galerkin-Method-based (NSE/FGM) flow solver for the incompressible Navier–Stokes equations, under the Boussinesq approximation in doubly non-periodic deformed domains \citep{diamantopoulos2022}. For the bathymetry of choice, two-dimensional simulations are conducted over a vertical plane normal to the isobaths. A third, Fourier-discretized, periodic direction may be readily incorporated in the transverse to conduct simulations aimed to resolve turbulence. Note that the bathymetry is one-dimensional (strictly a function of the along-wave direction), and any potential lateral variations in depth are not accounted for. Non-hydrostatic effects, a critically important part of shoaling ISW dynamics, are accounted for through a fast hybrid direct/iterative solver for the two-dimensional pressure Poisson equation (PPE) for each transverse Fourier mode based on a three-level Schur-complement approach based decomposition of the grid which is highly parallelizable (\citealt{karniadakis2005,joshi2016a,diamantopoulos2022}). Additional information about the required numerical stabilization of these inherently under-resolved simulations can be found in \citet{diamantopoulos2022} and Appendix Aa. Finally, the time-discretization scheme is based on \cite{karniadakis1991} and further explained in \cite{diamantopoulos2022}. The time step, $\Delta t$, is adaptively chosen during each simulation so as to respect the Courant–Friedrichs–Lewy constraints in all three directions.

\subsection{Simulation description} 
\label{subsec:Sim_descr}
\subsubsection{Initialization of the simulations} 
\label{subsubsec:initialize}
\par
All simulations are initially run in two-dimensional mode. The ISW is initialized on the deep-water artificial constant-depth region (Fig. \ref{fig:domain}a) and progressively advances into shallower waters across the entire 80 km-long propagation track. The term ``plateau'' will hereafter refer to the almost uniform-depth bathymetry that begins at the $68^{th} \:$km of the transect (see Fig. \ref{fig:domain}a). 

\par
The along-wave spectral element dimension is progressively decreased prior to ISW arrival at the region of convective instability, accordingly reducing the minimum grid-spacing, to ensure optimal resolution of the instability and resultant turbulence \citep{diamantopoulos2021}. Elements are finer within the unstable/turbulent ISW interior and coarser in the less active regions above/below it \citep{diamantopoulos2021}. 

\par
At least three initial wavelengths ($\simeq 3 $km) prior to the onset of the primary two-dimensional convective instability (see column $3D_{IN}$ in Tab. \ref{tab:ISW}  for exact location on the transect), the simulation switches to three-dimensional mode by extruding the domain in the periodic transverse direction. At this stage, finite-amplitude velocity and density perturbations, with characteristic velocities at 1\% of maximum wave-induced velocity field, are inserted (see details in Appendix Ab). A bypass transition to turbulence \citep{schmid2001}, through nonlinear interactions among different instability modes is thereby facilitated by the high amplitude of these modes. Turbulent flow structure is then found to occur well-before the arrival of the ISW at the deep mooring. 

\par In the study at hand, the wave propagation distance remaining from the location of perturbation insertion until the onset of convective instability ($ \sim \: 3 \: L_w$) enables any potentially laterally unstable mode or combination thereof, with a higher growth rate than the primary convective two-dimensional instability, to develop \citep{winters1992, winters1994}. Nonetheless, we have found that the primary convective instability in this study has always been strictly two-dimensional.

\par
Three production runs were conducted with different initial wave amplitudes ($\eta_{max}^{init}$ in meters): 136 ({\em {small}}), 147 ({\em {baseline}}), and 150 ({\em {large}}). The location of transition into a three-dimensional simulation for each wave is contingent on the critical depth of onset of the dominant two-dimensional convective instability, itself set by both the wave amplitude and the bathymetric changes, occurring earlier along the propagation track with increasing initial wave amplitude \citep{riverarosario2020}. All the ISW-associated deep-water characteristics, as well as the location of the associated three-dimensional initialization, are presented in Tab. \ref{tab:ISW}.

\begin{table}[t!]
    \begin{center}
        \begin{tabular}{|lcccccccc}
            \hline
            ISW & Amplitude & Wavelength & Celerity & Velocity & Wave Time Scale & 3D Initialization & \multicolumn{2}{c}{Non-Dimensional Characteristics} \\
            \cline{8-9}
            & $A$ (m) & $L_w$ (m) & $C$ (m/s) & $U_{max}$ (m/s) & $L_w / C $ (s) & $3D_{IN}$ (km) & $A / L_w$ & $U_{max}/ C$ \\
            \hline
            \textit{small}    & 136 & 1024 & 1.887 & 1.769 & 543 & 44   & 0.1329 & 0.9379 \\
            \textit{baseline} & 143 & 1015 & 1.926 & 1.843 & 527 & 40   & 0.1428 & 0.9568 \\
            \textit{large}    & 150 & 1010 & 1.957 & 1.898 & 516 & 31.5 & 0.1487 & 0.9695 \\
            \hline
        \end{tabular}
    \end{center}
    \caption{Deep-water characteristics and associated non-dimensional constants for each simulated ISW are presented. The position along the transect where each three-dimensional simulation begins is also indicated. Given the minimal variations in the deep-water wavelength among the ISWs, an approximate {\em {nominal}} wavelength of $1$ km (denoted as $L_W$) will be used  for normalizing the $x-$axis in each of Figs. 4, 6, 7, 8, 9, 10, 14 and 16f.}
    \label{tab:ISW}
\end{table}

\par The key differences with the predecessor simulations of \citet{riverarosario2022} in regards to the above may be summarized as:

\begin{enumerate}
\item Improved resolution in all three spatial directions, ranging within a factor of 2-to-3, which is enabled on the xz-plane through the use of variable-sized spectral elements (unlike the uniform-sized elements used in the predecessor study).
\item Application of spectral filtering (Sec. \ref{sec:Method}\ref{subsec:Num_Method}\ref{subsubsec:NSE_FGM}) on the xz-plane  only {\em {once}} (versus three times) per timestep, by virtue of a more robust numerical element-based discretization (spectral elements versus spectral multidomain penalty methods \citealt{joshi2016b}). When combined with a de-aliasing procedure (Appendix Aa) and the enhanced resolution above, the particular study offers non-trivially enhanced detail on the finer-scale structure of the instabilities and turbulence of interest.
\item  Insertion, at {\em {only}} the time of initialization of the three-dimensional simulations (versus in each overlapping window), of perturbations (Appendix Ab) \textit{O}(1000) stronger than those used by \cite{riverarosario2022}. As a consequence, per the enabled bypass transition to turbulence, the wave has a fully turbulent core prior to its arrival at the deep mooring. \cite{riverarosario2022} did not observe any three-dimensional instability structure before the shallow mooring and did not reproduce any Kelvin-Helmholtz instabilities in any stage of their simulations.
\item Due to the limited parallel performance of the flow solver used \citep{joshi2016c}, and the associated slower run turnaround, \cite{riverarosario2022} simulated only one ISW (corresponding to the baseline wave of this study) and not three.
\end{enumerate}

\subsubsection{Construction of the computational domain} 
\label{subsubsec:Comp_domain}
\par
\begin{figure}[h!]
 \centerline{\includegraphics[width=39pc]{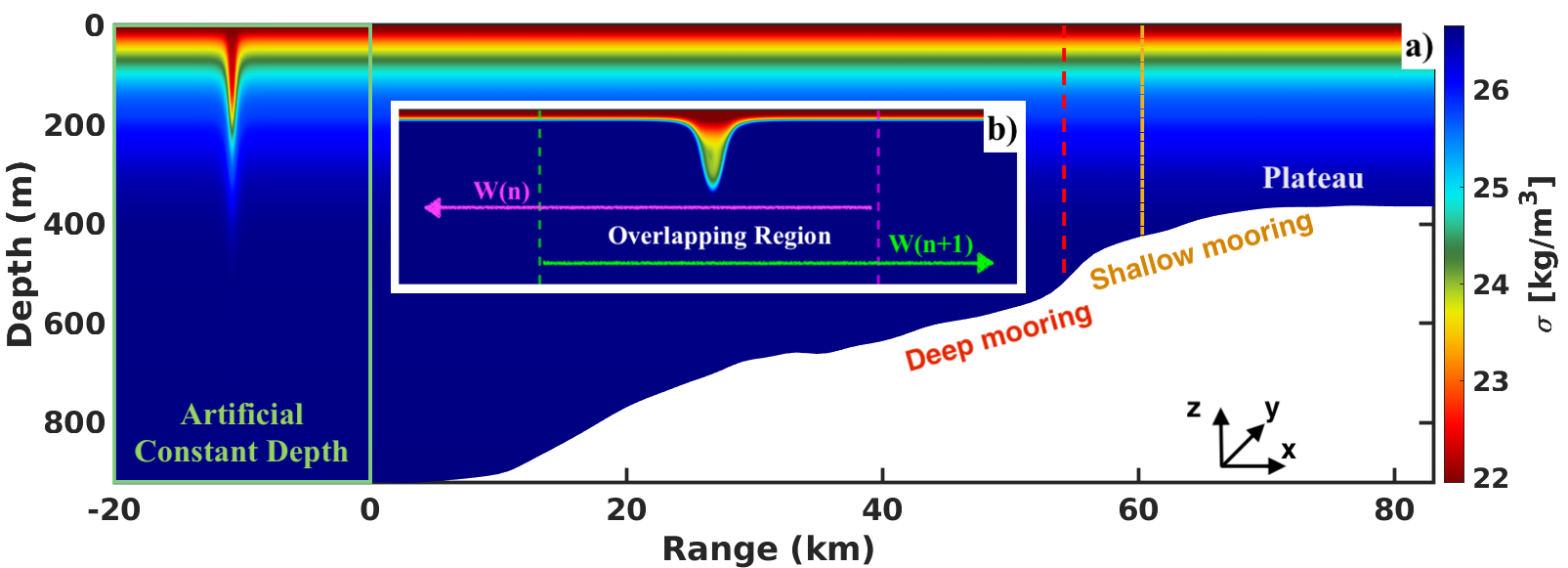}}
 \caption{(a): Full computational domain with density contours and initial ISW. Dashed vertical red and yellow lines indicate the locations of the {\em {shallow}} and {\em {deep}} moorings, respectively, as deployed in the SCS by \cite{lien2014}. (b): Exploded view of the ISW. The schematic emphasizes the overlapping region between two successive overlapping windows, denoted as $W_n$ and $W_{n+1}$ where $n$ is the window index. The approximate length of the overlapping region is 6~km, while the total length of each computational window spans 12~km.}
 \label{fig:domain}
\end{figure}

Figure \ref{fig:domain}a) shows adjusted density ($\sigma = \rho - 1000$ kg$/$m$^3$) contours in the full computational domain, with the initial ISW in deeper water. The bathymetric transect matches the one reported in Fig. \ref{fig:map}b (blue line). The locations of the original moorings deployed by \cite{lien2014} are also shown. Furthermore, the additional 20km artificial constant-depth region is denoted before the beginning of the transect (note the negative sign in Fig.'s \ref{fig:domain} horizontal axis). During the transition to the three-dimensional stage (see Tab. \ref{tab:ISW} for the exact location of each ISW), the computational domain is extruded in the y-direction with a width of $L_y \:=\: 50$ m. Per the rationale outlined in \cite{riverarosario2022}, the particular choice of the lateral domain width accommodates one wavelength of the fastest-growing transverse instability mode of the dominant two-dimensional convective instability \citep{diamantopoulos2022}.

\par
The simulations reported here are actually conducted in an overlapping window framework (see \cite{riverarosario2022} and \cite{diamantopoulos2022} for more detail): the computational domain is divided up into sub-regions successively overlapping in the direction of wave propagation. The three-dimensional window's length extends to approximately 12 wavelengths in the x-direction (corresponding to almost 12 km), while each overlapping region supports a propagation of 6 wavelengths (split between the two successive windows) on average, as illustrated via the exploded view near the ISW of Fig. \ref{fig:domain}b.  The simulation transitions to the next window along the shoaling track when the wave is at 3 wavelengths away from the right boundary of the current window. The use of the overlapping window allows one to efficiently simulate the ISW as a propagating feature localized, nevertheless, within a particular window. The maximum number of available grid points, as dictated by the computational resources at hand, is then focused on a single window. 

\par
Table \ref{tab:domain} summarizes the characteristics of the total domain and the associated three-dimensional computational window. Each three-dimensional overlapping window contains $46080$ spectral elements  on the x-z plane, each with a two-dimensional internal Legendre polynomial expansion of  order $p = 7$ \citep{kopriva} , which amounts to a total of $290$ million grid points per window. Per the flexible reduction in element size in both directions, as described in Sec. \ref{sec:Method}\ref{subsec:Sim_descr}\ref{subsubsec:initialize}, and the non-uniform spacing of the Gauss-Legendre-Lobatto grid points within each element \citep{canuto2007spectral}, particularly in the vertical direction inside the wave where elements are clustered, the highest resolution attained is summarized in Tab. \ref{tab:domain}. In the transverse direction, a Fourier-periodic grid with 128 points is employed, providing a resolution comparable to that in the \(x\)-\(z\) plane. Consequently, this study achieves significantly higher resolution in all three spatial directions ($\leq1$m) compared to \cite{riverarosario2022} (see also Appendix B). Additionally, soon after the onset of convective instability, $\Delta t$ is reduced through adaptive timestepping: $\Delta t$  remains consistently near-constant, however, between $0.1 \:$ and $\:0.2$ s across all runs. Finally, details on the Computational Resources used for these simulations of non-trivial computational cost are offered in Appendix Ac.

\begin{table}[t]
    \label{tab:domain}
    \begin{center}
        \begin{tabular}{lccc}  
            \hline
            Domain Length (km) & Window Length (km) & Domain Depth (m) & Domain Width (m) \\
            83 & 12 & [365, 921] & 50 \\
            \hline
            Polynomial Order ($p$) & Per window elements in $x$ ($m_x$) & Elements in $z$ ($m_z$) & Grid Nodes in $y$ ($N_y$)\\
            7 & 768 & 60 & 128 \\
            \hline        
            ${\Delta t}_{min}$ (s) & ${\Delta x}_{min}$ (m) & ${\Delta z}_{min}$ (m) & ${\Delta y}$ (m)\\
            0.1 & 1.03 & 0.2 & 0.4 \\
            \hline
        \end{tabular}
    \caption{The full computational domain, along with the associated overlapping window physical length scales, element count, and grid resolution, is shown. The finest resolution attained in each direction is also provided. Notably, each overlapping window utilizes all grid points in the \(z\) and \(y\) directions while focusing solely on a subset in the \(x\) direction.}
    \end{center}
\end{table}

\section{Convective Instability}
\label{sec:Conv_Evol}

\subsection{Baseline wave evolution: Flow structure}
\label{subsec:Baseline}
\par
\begin{figure}[H]
 \centering
 \includegraphics[width=39pc]{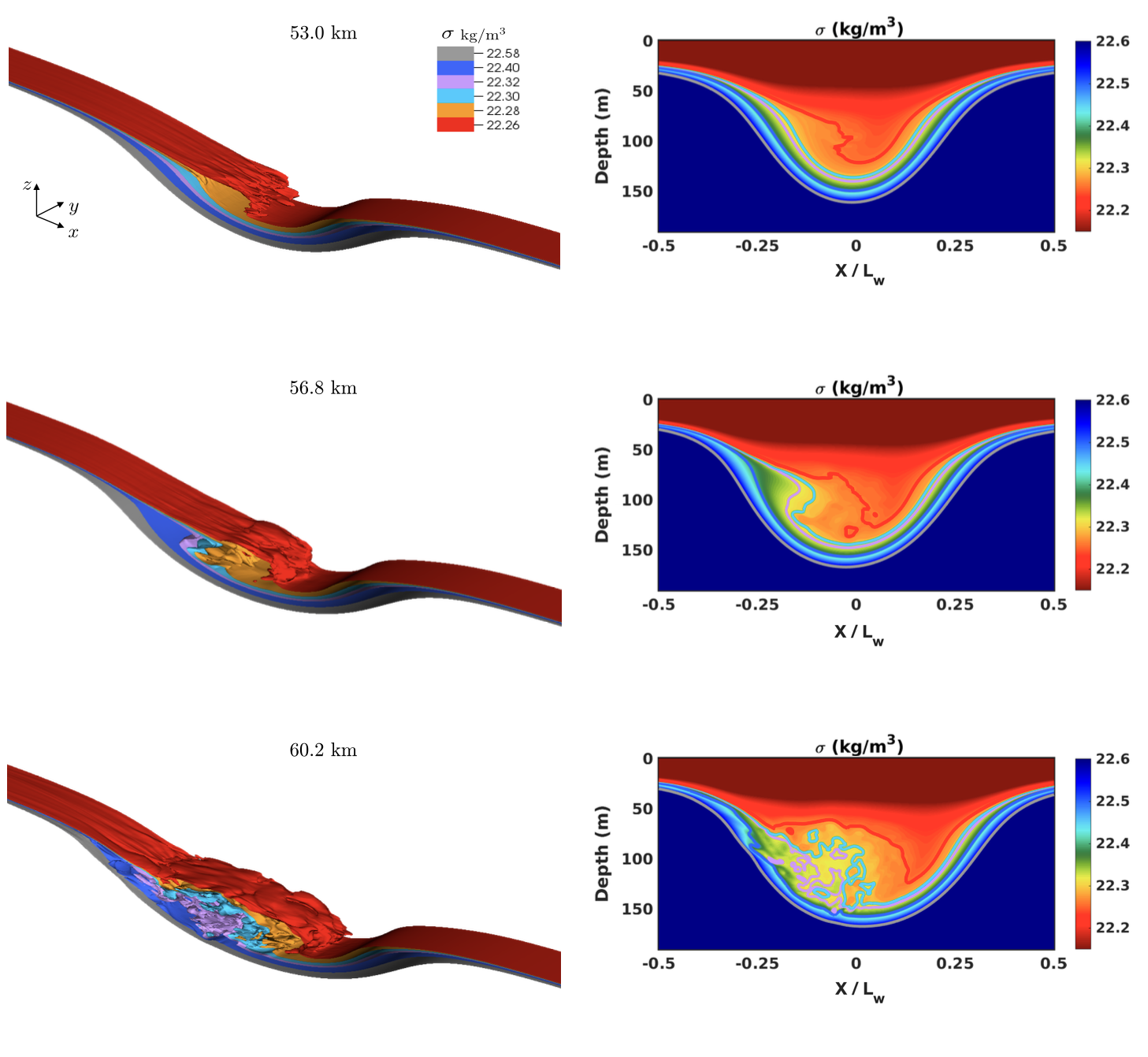}
 \caption{Evolution of three-dimensional convective instability in the baseline ISW ($A_{init}=143m$), shown via adjusted density ($\sigma = \rho - 1000$ kg$/$m$^3$) isosurfaces and contours. This quantity is used for all subsequent isopycnal surface/contour visualizations. \textbf{Left:} Three-dimensional $\sigma$ isosurfaces at 6 distinct values. \textbf{Right:} Mid-plane ($y=25$\,m) contours at the same downstream locations, overlapped with contour lines as a function of the normalized distance from the centerline with the common among the ISWs approximate wavelength ($\sim1$km); line colors and values correspond to the $\sigma$ isosurfaces of the same color in the left panels. Colorbar values are saturated accordingly to emphasize the convective instability and the entrainment and forward-plunging of fluid. The range of isopycnal surfaces and {\em {contour lines}} in Fig. \ref{fig:Baseline_Evolution_3D} is chosen such that its lower and higher bounds correspond to the first isosurface to convectively destabilize and to the peak in the background Brunt–Väisälä frequency in Fig. \ref{fig:insitu_profs}(d). A more dynamically evolving view of the convective instability may be found in the supplementary density animations and in \cite{diamantopoulos2021}.}
 \label{fig:Baseline_Evolution_3D}
\end{figure}

Figure \ref{fig:Baseline_Evolution_3D} shows the flow structure of the baseline ISW's core at three locations along the transect where convective instability dominates, by visualizing the adjusted density ($\sigma = \rho - 1000$ kg$/$m$^3$) field through three-dimensional isosurfaces (left) and contour plots sampled on the transverse midplane (right). Note that this quantity will be used for any isopycnal surface/contour visualizations hereafter  and will be referred to simply as ``density''. The reader is also recommended to consult the supplementary animations of density for a more dynamically evolving perspective on the evolution of convective instability. Finally, note that the variable in the horizontal axis, $X=x-x_c$, of the right panels of Fig. \ref{fig:Baseline_Evolution_3D} (and several subsequent figures) represents a local coordinate which consists of the position $x$ in a coordinate system fixed to the bathymetry (see Fig. \ref{fig:domain}) offset by the position of the ISW trough, $x_c$, along the shoaling track. 
\par
The range of isopycnal surfaces and {\em {contour lines}} in Fig. \ref{fig:Baseline_Evolution_3D} (and Fig. \ref{fig:Contours_All_Isws}) is intentionally chosen such that its lower and higher bounds correspond to two specific isosurfaces in Fig. \ref{fig:Baseline_Evolution_3D}. The first one, drawn in red, effectively illustrates the convective instability onset in the form of the first isopycnal that plunges into the ISW's interior. The second one, drawn in grey, is associated with the peak of the background Brunt–Väisälä frequency in Fig. \ref{fig:insitu_profs}(d) and is an indicator of the pycnocline. The former isosurface also serves as an approximate indicator of the leading edge and top interface of the horizontally advancing gravity-current-like feature discussed below (see also Figs. \ref{fig:Contours_All_Isws} and \ref{fig:PKE_RMS}).

\par
The initial stage of ISW convective instability is manifested through the steepening of the lightest visualized isosurface (red isosurface and isocontour in Fig. \ref{fig:Baseline_Evolution_3D}), which is accompanied by lateral finer-structure development at 53 km. As the wave shoals (and is even more evident in the supplementary density animations), progressively heavier isopycnals continue to plunge into its interior as colder water continues to be entrained into the wave. The relative contribution of advection and buoyancy in this process of entraining fluid into the ISW interior remains unclear.

The entrainment of heavier water into the inner wave region (curled cyan and purple iso-contour lines) is even more visible at 56.8 km in Fig. \ref{fig:Baseline_Evolution_3D}. As this entrainment process continues, a sufficient amount of heavier water accumulates inside the wave such that a horizontally-advancing gravity-current-like feature develops (see supplementary animations). Hereafter, this particular flow feature will be referred to as simply a ``gravity current''.

\par
When the simulated ISW arrive at the {\em {shallow}} mooring location (Fig. \ref{fig:Baseline_Evolution_3D} at 60.2 km), all detached isopycnals have overturned vertically and laterally due to the turbulence associated with the convective plunging and the gravity current. The gravity current, whose leading edge is denoted by the red isopycnal surface/contour, advances towards the ISW trough and, at later times, past it (see supplementary animations). As the ISW propagates up the slope, and gravity current turbulence continues to homogenize the entrapped fluid, the ISW maintains the convectively unstable condition of ($u_{max}/C > 1$). Determining quantitatively if the red isopycnal acts like a barrier between the ISW core and the shallower near-surface water or if there is any (and how much) turbulent entrainment across this interface requires further investigation, and is deferred to future study.

\par
The subtle growth, of $ \mathcal{O}(10~\mathrm{m}) $, in the vertical displacement of the dividing (gray) isopycnal in Fig. \ref{fig:Baseline_Evolution_3D} , as the wave moves into shallower waters (53-60km), indicates a growth in wave amplitude which is consistent with the evolution of this particular wave property reported in \citet{riverarosario2022}. At the same time, the waveform remains remarkably symmetric as indicated by the along-wave structure of the same isopycnal. Any steepening effect in the wave rear occurs for lighter isopycnals, internal to the wave, e.g. the red-colored one in Fig. \ref{fig:Baseline_Evolution_3D}. This preservation of the structural integrity of the wave, in the presence of a shallow/near-surface pycnocline, is in direct contrast to the simulations of \cite{vlasenko2005} where the pycnocline intersects the bathymetry and the unstable shoaling wave breaks and terminally disintegrates.
\par 
Over the same part of the transect, the wave appears to visibly broaden along with the above increase in amplitude. This finding appears to be consistent with the observations of \citet{duda2004} and predictions of strongly nonlinear theory \citep{helfrich2006,stastna2002,stastna2008}. Somewhat paradoxically, the calculation of ISW wavelength (see Appendix B) shows that it is reduced along the transect, in agreement with \cite{riverarosario2022}, a reduction which might be the result of normalizing with the wave amplitude which itself has grown (see above).

\subsection{Wave comparison: Convective Instability Condition Development and Flow Structure} 
\label{subsec:Conv_Br_All_ISWs}

\par
\begin{figure}[h!]
 \centerline{\includegraphics[width=39pc]{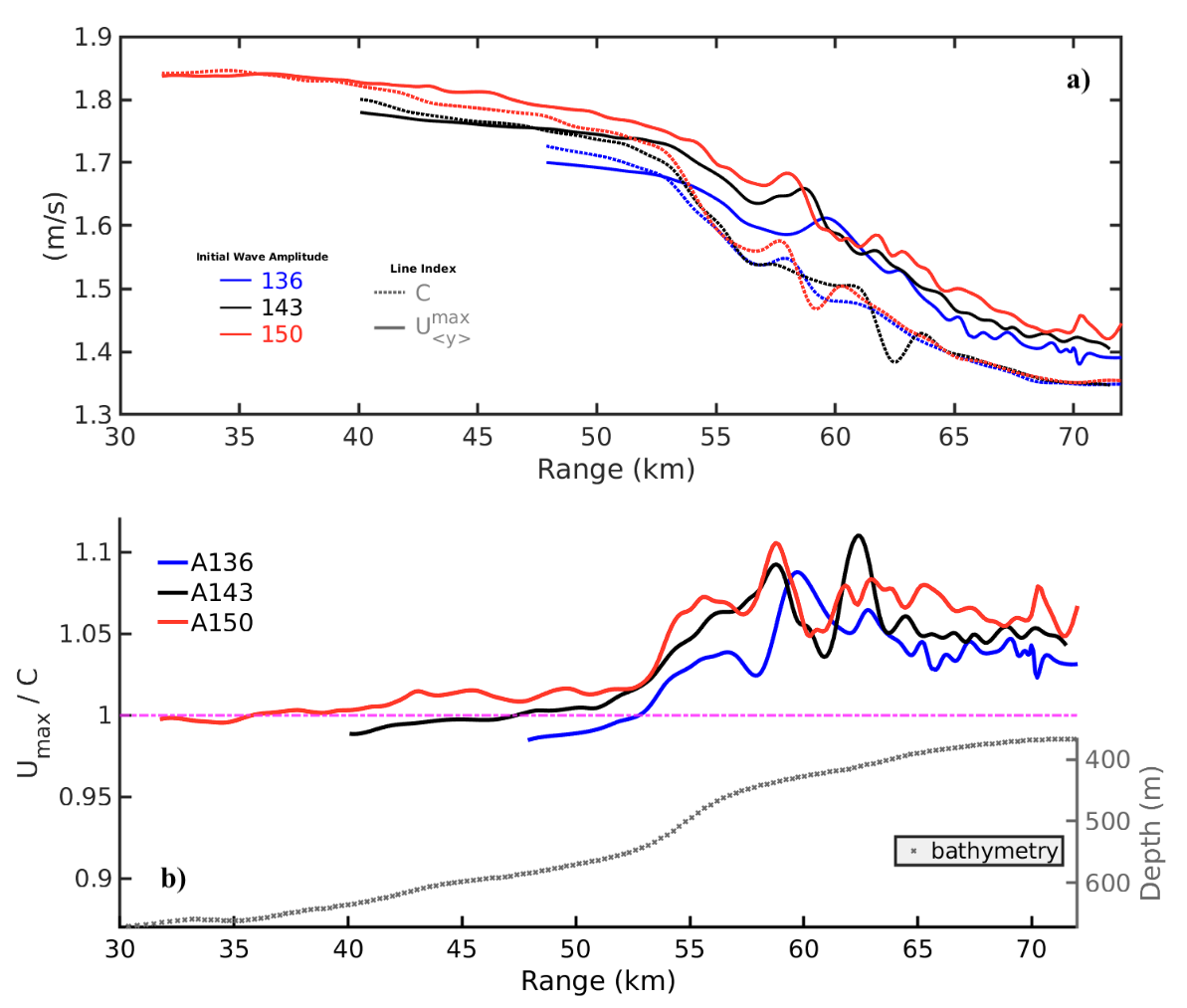}}
  \caption{(a) Maximum horizontal wave-induced velocity $u_{max}^{<y>}$ (solid lines) along with the propagation speed $C$ (dotted lines) are  for all 3 simulations as a function of the along-wave propagation range. (b) The corresponding ratio of $u_{max}^{<y>} / C$. The magenta dotted line (panel b) denotes $u_{max}^{<y>} / C = 1$. The gray markers in panel b) show the corresponding bathymetry. For each ISW, the corresponding curves begin at the location of initialization of the respective three-dimensional simulation (see Sec. \ref{sec:Method}\ref{subsec:Sim_descr}\ref{subsubsec:initialize} and Tab. \ref{tab:ISW}). Any indication of the potential for marginal convective instability is restricted to the {\em {large}} wave at the earlier stages of the wave evolution ($32-40 \: $km).}
 \label{fig:UmaxC}
\end{figure}

The convective instability develops in a similar manner across all three simulated waves. Figure \ref{fig:UmaxC}(a) illustrates the evolution of the maximum transverse-averaged horizontal wave-induced velocity, $u_{max}^{<y>}$, along with the corresponding propagation speed, $C$, for each wave. The propagation speed is estimated by computing the slope of a least-squares fit to the trajectory of the ISW trough, following the method described in \citet{chang2011} and \citet{riverarosario2020}. In Fig. \ref{fig:UmaxC}(b), the ratio of $u_{max}^{<y>}$, to the corresponding $C$, is also shown for each wave. The magenta dotted horizontal line marks the marginally unstable threshold where $u_{max}^{<y>} = C$, while key bathymetric changes are included for reference.  
\par
In the deeper part of the transect ($<$ 54 km), larger waves exhibit higher propagation speeds, consistent with the theoretical discussion of \citet{Stastna2022} (not explicitly shown). However, all three simulated ISWs follow a common trend of deceleration as they propagate over the slope region. The corresponding $u_{max}^{<y>}$ also decreases monotonically, albeit at a slower rate, eventually exceeding $C$. The critical depth for the onset of convective instability ($ u_{max}^{<y>} > C$) is clearly wave-amplitude dependent, occurring earlier along the transect for larger waves, in agreement with \citet{lamb2002} and \citet{riverarosario2020}. Convectively unstable conditions ( $u_{max}^{<y>} > C$ ) persist throughout the remainder of the ISW propagation, with $u_{max}^{<y>} / C$ increasing just before the waves reach the steepest slope region near the {\em{shallow}} mooring (58–60 km).  At this stage, by examining independently the evolution of $C$ and $u_{max}^{<y>}$ (Fig. \ref{fig:UmaxC}a), one first observes that all 3 waves decelerate at a nearly identical rate. In addition, a local maximum in $u_{max}^{<y>}$ occurs earlier with increasing wave amplitude at $[1.61, 1.66, 1.68]$ m/s, respectively. Shortly afterward, a second local maximum in $u_{max}^{<y>}$ is observed, as the deceleration continues to be synchronized among the three waves. Further along the transect, $u_{max}^{<y>}$ remains higher for larger-amplitude waves. However, all propagation speed curves effectively collapse together and reach a constant value of $1.35$m/s at $68$km once each wave has begun its ascent onto the plateau. Consequently, the actual ratio $u_{max}^{<y>} / C$ remains higher for larger ISWs. Finally, it is worth noting that the {\em{large}} ISW exhibits a propagation speed 42–53\% higher than that of a linear mode-1 long wave, as determined using a Taylor-Goldstein equation solver \citep{Stastna2022} configured with the background velocity/stratification profiles used here for depths ranging from 670 to 360 m.

\par
In regards to the {\em {large}} wave, marginal convective instability (a persistent state of $u_{max} \simeq C$; \citealt{chang2021a}) is potentially present per the proximity between the propagation speed and maximum current velocity curves between 32 to 42 km of the transect. Actually, during its propagation at $32$ km, the {\em {large}} ISW undergoes an early weak convective instability event: isopycnals initially detach from the rear but then return to their original position, until the subsequent, more powerful, isopycnal plunge takes place as linked to the dominant instability developing at 45 km (see supplementary animations). 

\begin{figure}[h!]
 \centering
 \includegraphics[width=39pc]{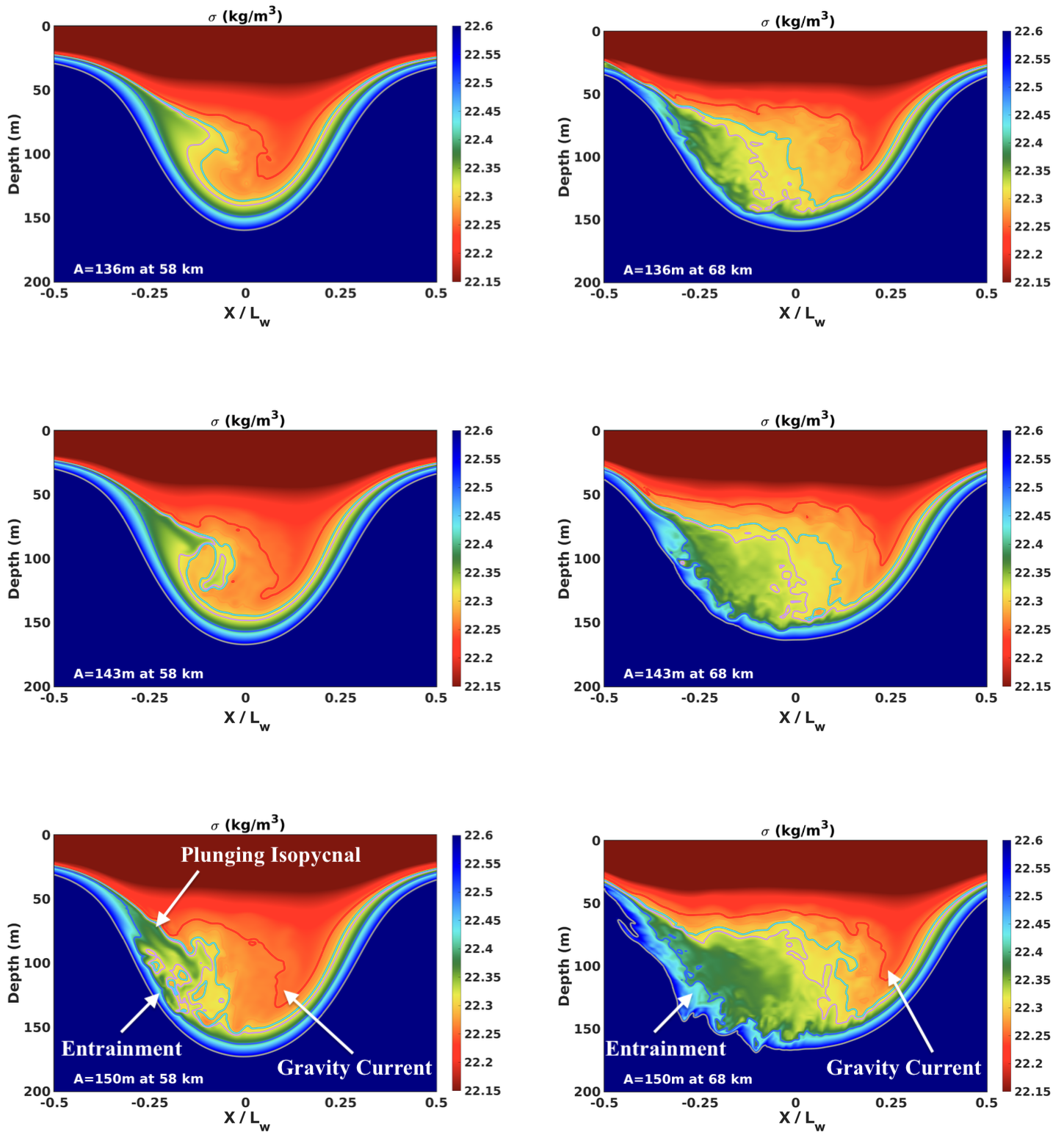}
 \caption{Mid-plane ($y=25$\,m) snapshots of the simulated waves captured during the convective instability and its subsequent evolution. \textbf{Left:} 58 km $\&$ \textbf{Right:} 68 km. \textbf{Top:} Small (A136), \textbf{Middle:} Baseline (A143), \textbf{Bottom:} Large (A150). The flooded density contours are shown, overlapped with the same 6 $\sigma$ contour lines of Fig. \ref{fig:Baseline_Evolution_3D}. The $x-$axis plots the distance from the centerline normalized with the nominal wavelength $L_W$ (see Tab. 1). This same $x$-axis normalization is used in several subsequent plots.}
 \label{fig:Contours_All_Isws}
\end{figure}

\par
Figure \ref{fig:Contours_All_Isws} shows mid-plane density contours zoomed in each wave's core region for the three cases (wave amplitude increasing from top to bottom). Snapshots are shown approximately at $58 \:$ km on the left panels, where the steepest slope lies, and at $68 \:$ km on the right panels where the simulated waves approach the almost-uniform-depth plateau. The ISWs appear to be broader and flatter with increasing amplitude, as also suggested by  \cite{lamb2002,stastna2002}. In the left panels, all ISWs are still in the initial forward plunging stage (see also Fig. \ref{fig:Baseline_Evolution_3D}). Yet, the larger the wave, the earlier the convective instability has occurred; more cold/heavy water has filled the wave interior. On the right panels, entrainment from the wave rear and mixing are even more visible with both effects being more vigorous for higher amplitude waves. Note that, at this stage, the well-mixed region established by the gravity current in the wave core (see Sec. \ref{sec:Conv_Evol}\ref{subsec:Baseline}) has advanced further towards the ISW's front edge; heavier water mass visibly occupies a greater volume of the ISW core for larger ISWs. This process continues until the end of the simulations, by which point the rear of the wave has been consistently more mixed with colder water with increasing ISW amplitude (see supplementary density animations).

\subsection{Lateral Structure} 
\label{subsec:Lateral_Struct}
\par
\begin{figure}[h!]
 \centering
 \includegraphics[width=39pc]{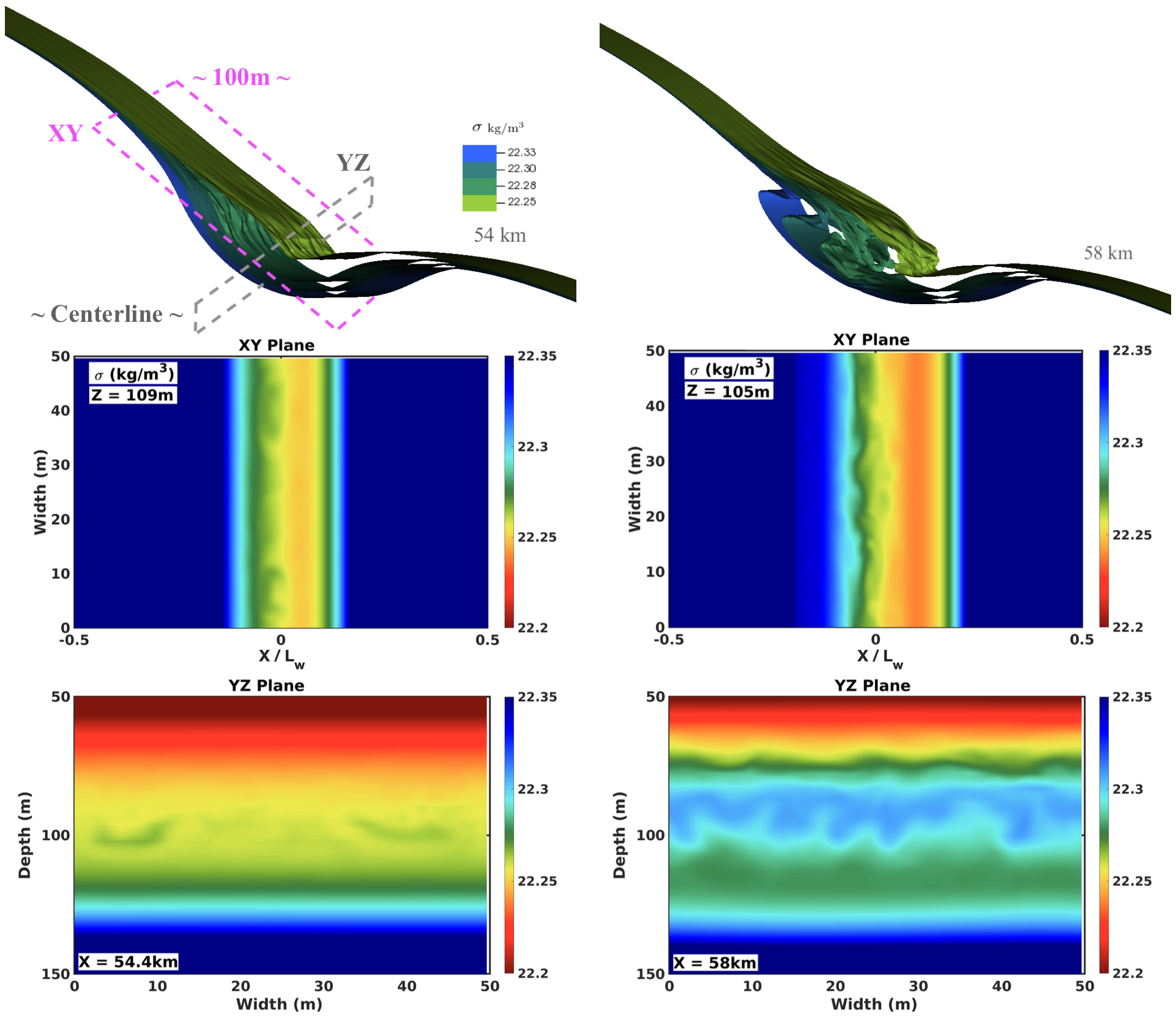}
 \caption{Spanwise structure of  the convective instability for the the {\em{small}} ISW. Two locations along the transect, approximately 54 km and 58 km (left and right panels, respectively), are shown. The top panels depict the three-dimensional structure using four selected isosurfaces, while also delineating the location of planar transects where contour plots are sampled in panels below. The middle panels show contours of density on $xy$ planar transects at approximately 100m depth. The bottom panels show equivalent $yz$ planar transects at ISW trough.}
 \label{fig:Lateral}
\end{figure}

Figure \ref{fig:Lateral} provides perspective on the transverse structure of the convective instability development by visualizing the density field. The presentation is restricted to the {\em{small}} ISW  where any spanwise features are distinctly identifiable. The two larger-amplitude waves involve a  more complex lateral structure, a direct result of the bypass transition procedure in three-dimensional simulation initialization (see Sec. \ref{sec:Method}\ref{subsec:Sim_descr}\ref{subsubsec:initialize}). Visualizations are shown in Fig. \ref{fig:Lateral} at two locations along the transect: the first is situated just after the onset of convective instability (left panels, at \(\sim 54 \, \text{km}\); cf. the blue curve in Fig. \ref{fig:UmaxC}) and the second is near the steepest slope (right panels,  at \(\sim 58 \, \text{km}\)), immediately before the first vigorous plunging event occurs (see top left panel of  \ref{fig:Contours_All_Isws} and supplementary animations of density).  The top panels of  Fig. \ref{fig:Lateral} show select three-dimensional isosurfaces of density. The middle and bottom panels of the same figure show density contours on specific stream-span  ($xy$)  and span-depth ($yz$) planar transects at approximately 100m depth and the ISW trough, respectively. Note that the density differences within the flow features of interest are subtle, requiring the limited colorbar value range in Fig. \ref{fig:Lateral} to effectively visualize the particular features in the density field. It is the high-order accuracy spatial discretization which provides confidence in that these fine variations are physically accurate. 

\par
During the onset of convective instability (left panels), two intruding lobes (laterally offset from the $y$-mid-plane) can be identified on the \(\sigma = 22.28\) kg/m$^3$ isopycnal contour,  in the middle panel ($xy$-transect). A set of two weaker intruding lobes, symmetrically positioned around the $y$-mid-plane, is also visible. The two laterally offset lobes are associated with weak small-scale lateral overturns in the lower panel ($yz$-transect). Visualizations of isosurfaces and equivalently sampled contours on planar transects of the transverse $v$-velocity (not shown here; a visualization tool commonly used to examine three-dimensional instabilities; \citealt{sakai:20a}) show a consistent spatial structure. Per the bypass-transition strategy used, where perturbations are inserted with uniform energy content across the first 7 lateral Fourier modes, the absence of a distinctly sinusoidal larger-scale transverse oscillation in the isopycnal contour of interest indicates that it is not a single lateral instability mode/wavelength operative at this location of the shoaling track. Instead, a more complex lateral instability structure is at play, involving the interaction (at weak, yet finite amplitude) of the longer lateral wavelengths the computational domain width can accommodate. 

\par 
In the right panels (58km), the entrained and convectively unstable density isosurfaces propagate toward the front of the ISW while overturning, giving rise to a less organized transverse flow. A shorter-scale undulation is observed on the \(\sigma = 22.28 \) kg/m$^3$ isopycnal contour on the $xy$-transect,  presumably linked to a similar feature at $75$m depth in the $yz$-transect. This latter transect also shows a finer-scale feature extending over a depth range of $[85,105]$m for colder water at values of \(\sigma=22.30\) kg/m$^3$.

\par Similar flow features are observed in the two larger simulated waves (not shown here). The lobe-like or oscillatory structure of the equivalent isopycnals at the onset of convective instability is subject to additional non-trivial distortions, with more vigorous overturns visible on $yz$-planar transects even before the gravity current (see Sec. \ref{sec:Conv_Evol}\ref{subsec:Gravity_Current}) is established within the ISW interior. 

\par Finally, we emphasize that the computational domain’s transverse dimension (its choice dictated in part by limitations in available computational resources) might not be wide enough, in the sense that it may restrict the development of convective instability transverse length-scales which may be larger than the model domain width itself. A quantitative study of the role of computational domain width in this regard, in the spirit of what has been done for turbulence transition in separating boundary layers in aerodynamics \citep{zhang:16}, is a non-trivially ambitious exercise which is deferred to future investigations.

\subsection{Background Recirculation} 
\label{subsec:Background_Recirc}

\par
\begin{figure}[h!]
 \centering
 \includegraphics[width=39pc]{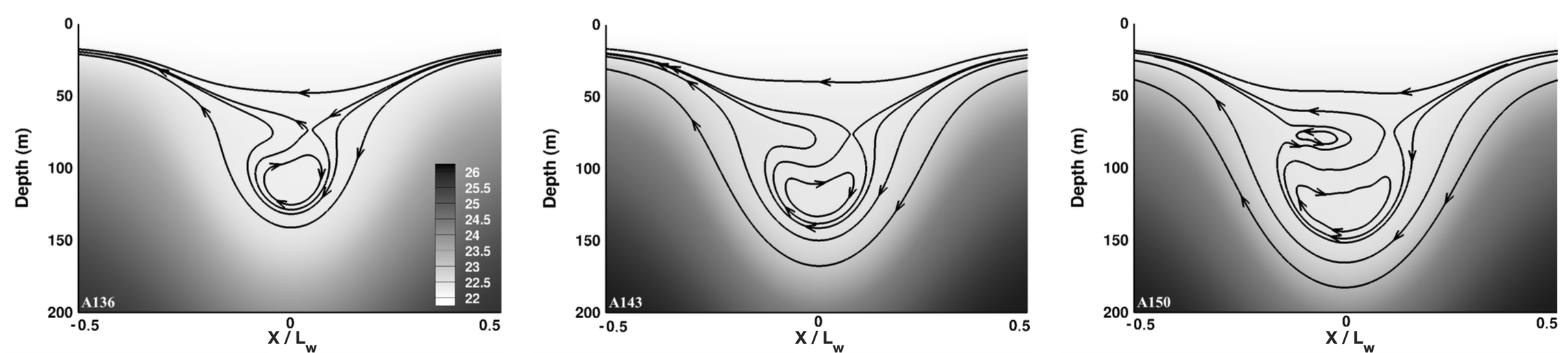}
 \caption{Velocity streamlines overlaid on density contours $\sigma$. The arrows on each streamline indicate the direction of the flow. Both density values and flow direction correspond to the y-averaged field computed in a wave-attached frame of reference. The panels compare the three simulated waves at the {\em {deep}} mooring location, with increasing amplitude from left to right. A distinct large region with closed streamlines is located in the center of the wave core for all three ISWs, while a second one smaller in size is formed in the {\em {large}} wave, creating a pair of counter-rotating vortices. The choice of colormap differs from the one used in previous figures to better highlight the streamlines. In all panels, $x-$axis is the normalized distance from the wave centerline (see Tab. 1).}
 \label{fig:Streamlines}
\end{figure}

Streamline visualization is used to investigate the potential for a two-dimensional background subsurface recirculation (see Sec. \ref{sec:intro}) to form inside the turbulent core of the simulated convectively unstable ISWs, per the equivalent observations of \cite{lien2012}. Figure \ref{fig:Streamlines} shows streamlines produced by the transverse-averaged velocity field, overlaid on density contours. The panels show a distinct background {\em {subsurface}} recirculating flow with closed streamlines for all ISWs respectively, sampled at the location of \cite{lien2012}'s in-situ observations. This primary subsurface recirculation (located in the trough of each wave) is found to occur earlier along the transect with increasing wave amplitude (not shown), and is associated with the ISW arriving at a critical depth where the convective instability condition, $u_{max}=C$, is satisfied. Moreover, the streamline structure of Fig. \ref{fig:Streamlines} is similar to the two-dimensional simulations of \cite{riverarosario2020}. However, in this previous study,  closed-streamline formation is delayed and is not observed, for the wave equivalent to our baseline case, until the wave arrives at the {\em {shallow}} mooring location. 

\par
In the simulated {\em {large}} ISW, a smaller-scale secondary closed-streamline recirculating region develops, associated with the plunging isopycnals in Fig. \ref{fig:Contours_All_Isws} and baroclinic vorticity generation therein. Consequently, a pair of counter-rotating vortices is established, offering further support for the  observations of \cite{lien2012} which assumed steady-state. In analogy with the time of emergence of the primary subsurface recirculation, the formation of this secondary vortex structure is delayed with decreasing wave amplitude. As the ISWs advance to shallower water, the two aforementioned counter-rotating vortices interact and move around the rear of the wave. Further downstream, smaller-scale vortical motions develop inside the wave core as the inner region of the ISWs transitions to a less organized state.

\par By observing the structural response of the isopycnal field in the rear of the wave interior where the above counter-rotating vortex pair develops (see the supplementary density animations), it does appear that this vortex pair is one way through which, at least initially, entrainment of heavier fluid into the ISW core is facilitated. Future analysis, relying on the Lagrangian tracking \citep{klose2020} of actual particles entrained into the ISW interior, would be needed to substantiate this observation.

\subsection{Gravity Current} 
\label{subsec:Gravity_Current}
\par

\par
\begin{figure}[h!]
 \centering
 \includegraphics[width=39pc]{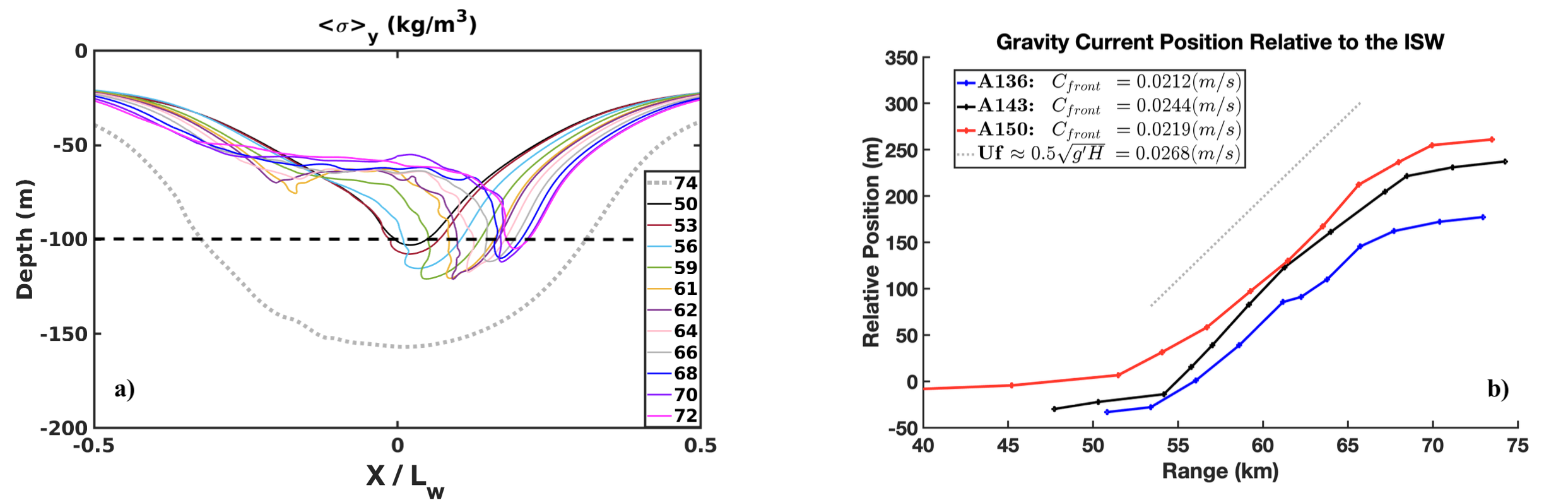}
 \caption{Solid-colored contour lines in panel (a) represent the $<\sigma>_y = 22.26 \:$ kg/m$^3$ isopycnal (value that corresponds to red isopycnal in Figs. \ref{fig:Contours_All_Isws} and \ref{fig:Ri}) of the {\em{small}} ISW. The ISW centerline position for each sample is indicated in the legend (in km). The grey dotted line represents the  $<\sigma>_y = 22.58 $ kg/m$^3$ isopycnal (grey isopycnal in Figs. \ref{fig:Baseline_Evolution_3D}, \ref{fig:Contours_All_Isws}, and \ref{fig:Ri}), effectively denoting the pycnocline, sampled at the final snapshot of the simulation. The grey horizontal dashed line at 100 m depth highlights the sampling location for the horizontal gradient of the density field (see main text). Panel (b) depicts the along-wave position of the approximate gravity current front for each ISW. The grey line represents the theoretical prediction for front position in idealized lab-scale lock-exchange gravity current experiment, subject to a density jump of $\delta \rho = 0.1 \:$ kg/m$^3$ in tank of depth $H = 3$m, for reference (see in-text discussion). The corresponding front propagation velocities are shown in the legend.}
 \label{fig:Grav_Curr}
\end{figure}

  
The gravity current discussed in Sec. \ref{sec:Conv_Evol}\ref{subsec:Baseline} and \ref{sec:Conv_Evol}\ref{subsec:Conv_Br_All_ISWs} propagates toward the front of the wave in a similar manner across all simulated ISWs. In a reference frame moving with each ISW, the transverse-averaged isopycnal \(\langle \sigma \rangle_y = 22.26 \) kg/m$^3$ (corresponding to the same value as the red isopycnal in Figs. \ref{fig:Contours_All_Isws} and \ref{fig:Ri}) is used as an indicator of the gravity current front and is tracked across multiple instances during the up-slope wave propagation. Figure \ref{fig:Grav_Curr} (a) illustrates this tracking procedure for the {\em{small}} wave where the isopycnal of interest is shown with a solid line, effectively illustrating the advancement of the vigorous gravity current event toward the front of the ISW. To this end, the transverse-averaged isopycnal corresponding to the pycnocline is also shown with the gray dotted line for the final snapshot of the simulation. The corresponding ISW trough locations along the transect are indicated in the legend (in km). At approximately $100$m depth (denoted by the grey dashed line in panel (a), the gravity-current-front-tracking isopycnal is nearly vertical. At the particular depth, and at each location examined along the shoaling track, the along-wave position of this vertically-oriented isopycnal is then graphically estimated and used to calculate its distance from the ISW centerline.  This procedure has been followed for all three simulated waves and the evolution of the along-wave offset of the particular isopycnal is plotted in Fig. \ref{fig:Grav_Curr} (b) as a function of the corresponding ISW trough location along the transect. 

\par  
The speed at which the gravity-current-front-tracking isopycnal advances from the rear to the front of the ISW is estimated through a least-squares straight-line fit of its relative position to the centerline for a focused range of ISW trough locations along the shoaling track. For all three simulated waves, this range of locations is restricted to the interval $[55,65]$km, where the gravity current front appears to propagate at a near-constant speed and the convective instability is at its most energetic state (see Fig. \ref{fig:UmaxC} and supplementary animations of density).  The estimated speeds are listed in the plot legend. Visual inspection of the gravity current's frontal position in Fig. \ref{fig:Grav_Curr} (a) indicates that it propagates toward the front of the wave at a comparable speed of $0.02$m/s {\em{ across all three simulated waves}}.   

\par Closer examination of the supplementary animations of density suggests that, although the convective instability is triggered earlier along the shoaling track with increasing wave amplitude, the associated entrainment of heavier/colder water appears to initiate from the depth of the \(\langle \sigma \rangle_y = 22.26 \) kg/m$^3$ regardless of deep-water wave amplitude.  Moreover, the plunging of this entrained water, and resulting horizontal acceleration of the gravity current within the ISW, appear to be facilitated by the more rapid bathymetric change near the steepest slope (at around $54$km along the shoaling track) and the subsequent deceleration of the propagating ISW. These commonly shared effects may be why the gravity current propagation speed is constant across all three waves. As also stated in the previous section, future work with Lagrangian-particle-tracking techniques \citep{klose2020} would help justify this interpretation.

\par The horizontal variation of the density field across the gravity-current-front along the 100 m-depth dashed line (Fig. \ref{fig:Grav_Curr}a) has been found to linearly between $22.25$ and $22.35$~$kg/m^3$ for all three waves. If one uses idealized two-layer lock-exchange theory \citep{linden2012} with this difference of $\delta \rho = 0.1 kg/m^3$, an effective water tank depth of $H=3$m would be needed to produce a gravity current propagating at a comparable speed of $0.0268$m/s (see the dotted line in Fig. \ref{fig:Grav_Curr}b). Two-layer lock-exchange theory is effectively not applicable here, as the height of the gravity current advancing within the simulated ISWs has a  vertical extent which at least one order of magnitude larger.  Existing studies of gravity currents splitting off downward sloping boundaries, as encountered over the oceanic continental slope and in lakes/reservoirs, into a deeper stratified water column and modeled in the laboratory and through direct numerical simulations \citep{cortes:14,marques:17} may provide a theoretical framework to more accurately model the propagation the gravity currents discussed here. 

\par It is worth contrasting how markedly smaller the above gravity current propagation speed of $0.02$cm/s is when compared to that of density currents associated with oceanic overflows where the corresponding speeds are a factor of 25-to-50 larger \citep{legg:09} despite across-front density variations that are comparable to the value of $\delta \rho =0.1~kg/m^3$ reported above.  However, in the flow under studied here, this density variation occurs over a distance of $100$m which may responsible for this significantly slower front advancement velocity. The role of the convective-instability/turbulence-driven velocity fields within the wave also merits additional further study in this context. 

\par The relative gravity current front position is plotted in Fig. \ref{fig:Grav_Curr} (b). By the time the ISWs reach the plateau, and the gravity current has ``felt'' the front of the ISW, this distance of the front from the wave centerline is approximately $180, \: 240$ and $260$m, with increasing ISW amplitude. This slow advancement of the gravity current front  may be contrasted to the approximately $20$km traveled by the wave during a physical time slightly above $4$ hours. 

\par The relative position of the gravity current front is also a measure of the resulting volume of the wave's core that both the entrained and, subsequently, mixed fluid occupies.  Consequently, larger ISWs allow the gravity current to penetrate further into their interior and towards their leading edge, occupying a greater volume where finer-scale structures dominate and are expected to cumulatively drive even stronger mixing within the wave and its base during the above slow advancement. This observation aligns with the increased kinetic energy associated with velocity perturbations in higher-amplitude waves, as discussed in Sec. \ref{sec:Kinetic_Energy}. 

\par
Finally, using the pycnocline as a robust delineator of the ISW waveform in the simulations, the waveform remains fairly symmetric for all waves at least until they reach the plateau. To this end, although the gravity current plays a complex role in flow kinematics in the interior of the ISW, its impact on larger/wave-scale kinematics appears to be minimal.  

\section{Shear Instability} 
\label{sec:Shear_Inst}

\subsection{Richardson number} 
\label{subsec:Ri}

\par
\begin{figure}[h!]
\centering
 \includegraphics[width=39pc]{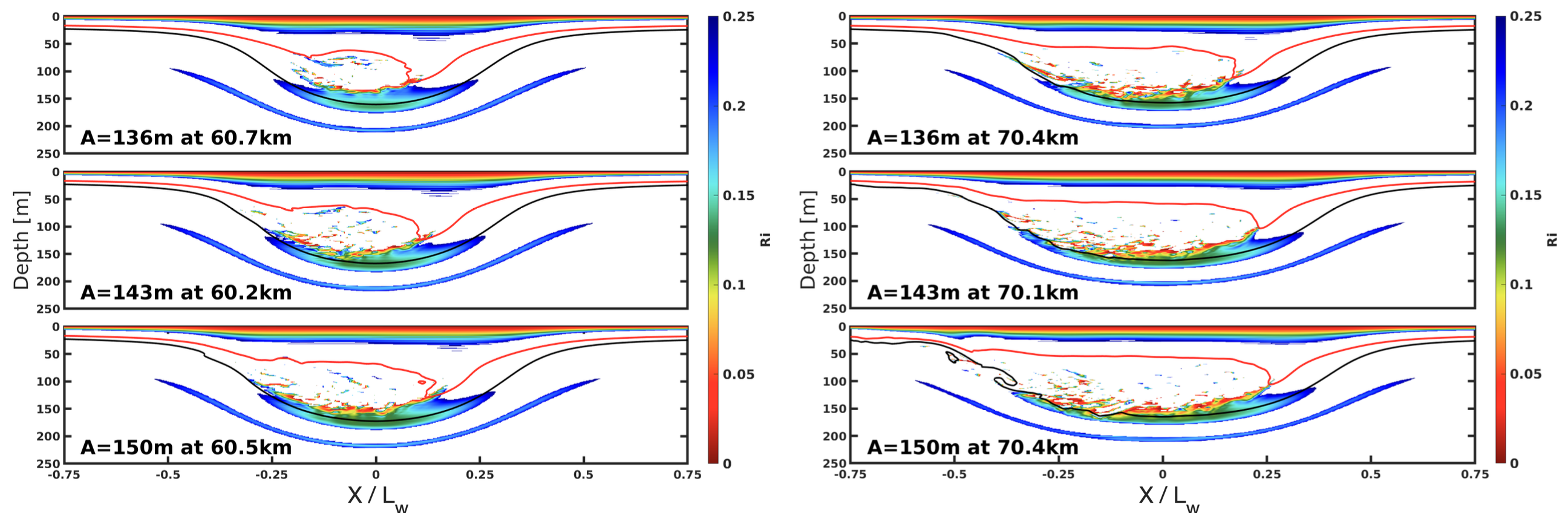}
 \caption{Transverse-averaged Richardson number contours for all waves at two locations on the shoaling track, at 60 and 70 km, respectively. Values only in the range of $[0, 0.25]$,conditions which are considered favorable for shear instability, are plotted. The red and black contour lines correspond to those in red and gray in Figs. \ref{fig:Baseline_Evolution_3D} and \ref{fig:Contours_All_Isws} and are used to delineate the subsurface recirculating core and the ISW waveform. The $x-$axis represents the distance from the centerline normalized with the approximate wavelength, which is common for all waves.}
 \label{fig:Ri} 
\end{figure}

The Richardson number ($Ri=N^2 / S^2$) may be used as a diagnostic for shear instability in the simulated ISWs (see Sec. \ref{sec:intro}). Here, the background stratification ($N$) and shear ($S$) are computed using the vertical gradients of the y-averaged fields of density and u-velocity. The critical value for shear instabilities to occur is well-known to be $Ri_c^{p} = 0.25$ for parallel stratified shear flows \citep{Smyth_Carpenter_2019}. For the finite-length curved stratified shear layer within the ISWs, the adapted criterion of $Ri_c^{c} = 0.1$ has been proposed (\citealt{fructus2009,barad2010,carr2011,chang2021b,Passaggia_2018,Xu2019}). The alternative criterion of a critical value of $L_x/L_w$ ($L_x$ and $L_w$ are the length scale of the region with $Ri < 0.25$ and the ISW wavelength) \citep{fructus2009,carr2011} is not examined in the present study due to the absence of a spatially-coherent-enough band of subcritical Richardson numbers needed to extract $L_x$ (see Fig. \ref{fig:Ri}).

\par 
 In Fig. \ref{fig:Ri}, the distribution of the computed $Ri$ number is shown for all three waves, for two locations on the transect: at the {\em {shallow}} mooring (left panel) and at the plateau (right panel). The colorbar limit is set to $[0,Ri_c^{p}]$ to emphasize the potential for shear instability in a stably stratified background. Inside each simulated ISW core, yellow-to-red colored contours, corresponding to $Ri \: \leq \:0.1$ criterion are found to extend from the ISW trough to its rear and are weakly offset above the pycnocline (black contour line). These regions of subcritical $Ri$ have formed due to the mixing and associated weakening of the local stratification produced by the convective instability, and its subsequent evolution, as discussed in Sec. \ref{sec:Conv_Evol}\ref{subsec:Baseline} and \ref{subsec:Conv_Br_All_ISWs}.

\par
The left panels of Fig. \ref{fig:Ri} display scattered yellow and red patches, while in the right panels (sampled at the plateau), a thin (yet distinct) layer carries a higher concentration of yellow-to-red regions which extend more contiguously over a larger region of the wave. This layer is longer and thicker and has enhanced spatial coherence with increasing ISW amplitude.

\par Note that this contiguous layer of lower $Ri$ at the plateau is directly connected to a density profile that is very well-mixed at the base of the wave around its centerline, with an underlying sharper stratification, as contrasted to the deep-water wave (not shown here). At this location on the shoaling track, the convective-instability-driven gravity current has propagated past the middle of the wave (Fig. \ref{fig:Grav_Curr}a) and has established a well-mixed region above the pycnocline. Moreover, as indicated by Fig. \ref{fig:Grav_Curr}b and the discussion in Sec. \ref{sec:Conv_Evol}\ref{subsec:Gravity_Current}, the gravity current extends further to the right of the wave centerline with increasing wave-amplitude yielding a broader well-mixed region of lower $Ri$.

\par Lastly, the robust yellow/red band displayed at the surface region (upper $ 20 \: $m) is formed due to the background current shear and is not examined in this study. The lack of an inflection point in the linearly extrapolated near-surface background velocity profile precludes the formation of shear instability therein. Moreover, the absence of any inserted perturbations during three-dimensional initialization within this narrow near-surface layer (Sec. \ref{sec:Method}\ref{subsec:Sim_descr}\ref{subsubsec:initialize}) and the non-trivial vertical offset of the subsurface phenomena of interest do not allow for any shear-driven turbulent production within it \citep{jacobitz1997}.

\subsection{Kelvin-Helmholtz instability development} 
\label{subsec:KH_Inst}
\par

In situ observations of \cite{moum2003} in ISWs at the Oregon shelf report $10$m-high K-H billows. Subsequent numerical investigations examined the occurrence of shear instability in two-dimensional simulations of ISWs, both in uniform-depth water \citep{barad2010} and shoaling bathymetry \citep{lamb2011}, reporting a K-H billow thickness of $10$m and $5$m respectively; the former study also involved a focused study of three-dimensional effects in billow development. In the SCS, the observations of \cite{lien2014} suggest the existence of shear instabilities, per the measurement of values of Richardson number, $Ri<0.11$. On the continental shelf of the same region, \cite{chang2021b} actually recorded shear-instability-driven roll-up thermal patterns with vertical scale varying from $15$m to $30$m in ISWs.

\par
\begin{figure}[h!]
\centering
  \includegraphics[width=39pc]{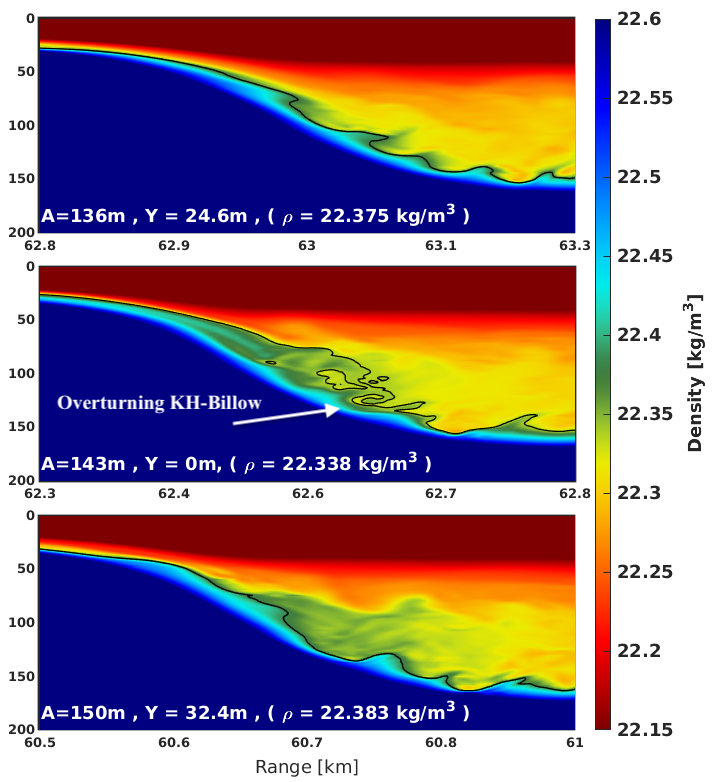}
  \caption{Shear-induced roll-up patterns in the density field within the ISW rear, captured within all three waves at approximately 60 km along the transect. The associated $x$ coordinate of the wave trough, the $y$-location of the selected stream-depth transect, and the reference density (black) contour line value ($\rho$) for each wave have been chosen optimally to best illustrate the roll-up patterns.}
  \label{fig:Early_KH}
\end{figure}

 Figure \ref{fig:Early_KH} shows density contours which demonstrate the earlier occurrences of shear-instability-induced roll-up patterns for all three simulated ISWs at the vicinity of the {\em {shallow}} mooring ($60\:$km). With the exception of a very distinct K-H billow in the {\em {baseline}} wave (middle panel of Fig.\ref{fig:Early_KH}) , the remaining roll-up patterns in all waves are not shown to fully overturn, at least yet. Whether overturning has not taken place yet, because the shear instability is still developing or it lacks in kinetic energy to overcome the local stratification, requires further investigation. To this end, a higher postprocessing file output rate than the one reported in Appendix Ac) would be highly advantageous. The less contiguous spatial distribution of subcritical values of the local Richardson number along the ISW trough at this location yellow-to-red band in the left panels of Fig. \ref{fig:Ri} would also need to be taken into account. Even without full overturns, evidence of K-H-like patterns that are still shown to form could potentially support \cite{chang2021c}, who suggests that K-H billows could potentially occur for ISWs with measured values of $0.1 < Ri < 0.25$. Therefore, concluding whether the proposed $Ri_c^{c}$ criterion for ISWs is too restrictive needs further investigation. Similarly, whether the roll-up patterns' puffy structure implies a Holmboe instability, particularly of the Asymmetric type, \citep{carpenter2007,carr2017,salehipour2016,salehipour2018,Olsthoorn2023} is deferred to future study. 

We reiterate that \cite{riverarosario2022} did not observe any clear pattern of shear instability due to the combination of limited resolution and insertion of weaker three-dimensional perturbations (Sec. 2 e 1). In agreement with \cite{chang2021b}, shear-instability-driven overturning features of Fig. \ref{fig:Early_KH} have a maximum vertical extent of $20-30$m in all three waves, as calculated according to \cite{carr2017}. Transverse-averaged density and along-wave-velocity profiles sampled at the centerline of the wave trough (not shown), show a 100m thick shear layer associated with the ISW signature, equivalent to the measurements of \cite{lien2014}. However, the K-H-billows of Fig. \ref{fig:Early_KH} are linked directly with the upper part of this shear layer, where the lower Ri number region is also found (between 100 and 140m meters, Fig. \ref{fig:Ri}). Lastly, at this stage and across all waves, the K-H-like roll-up patterns are correlated with sufficiently low local {\em{Ri}} values. Moreover, they are observed to be localized in the transverse direction (not shown) suggesting a potential similarity with the work of  \citet{Smyth2004} the exploration of which is, nonetheless, outside of the scope of this study.

\par
At later times, as the ISW propagates over the plateau, K-H billows with a complete overturn (not shown) are observed in all simulated waves. For the {\em {small}} and {\em {baseline}} waves, as shown in the supplementary animations, intermittent pulses of fully-overturned K-H billows are released towards the ISW tails. With the exception of a couple of isolated events, these particular billows mainly form on lighter isopycnals ($\sigma = 22.3-22.4$ kg/m$^3$) than the pycnocline, similar to those delineated in Fig. \ref{fig:Early_KH}. Their height is not significantly greater than what is reported for the events in Fig. \ref{fig:Early_KH}.

\par
\begin{sidewaysfigure}[h!]
 \centering
 \includegraphics[width=50pc]{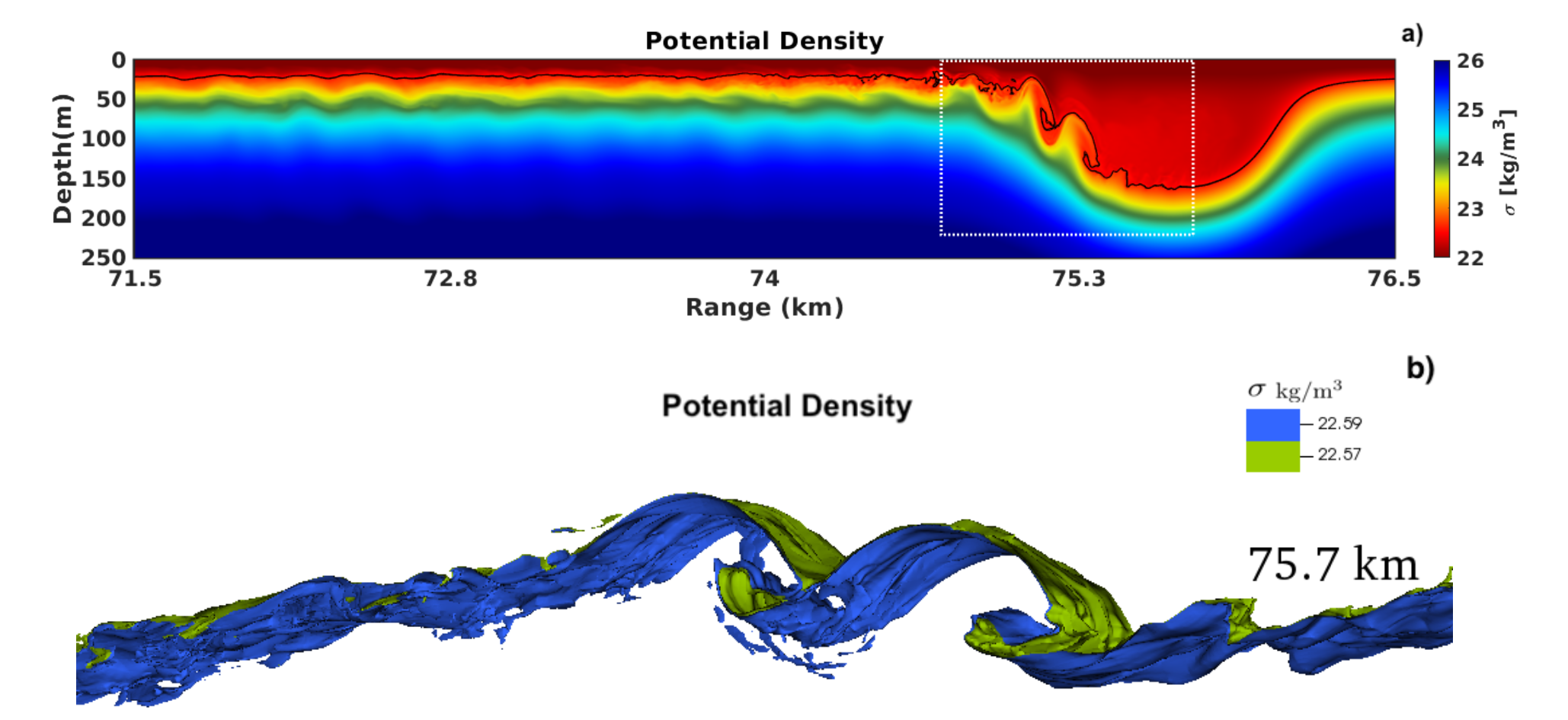}
\caption{Instantaneous snapshots of large K-H-billows forming in the {\em {large}} ISW rear captured on the plateau at a wave trough location of 76 km: two-dimensional transect of density contours ($a$) and three-dimensional isosurfaces ($b$). The three-dimensional snapshot ($b$) is sampled over the white-outlined region of ($a$) and includes two $\sigma$ isosurfaces (displayed as black isopycnals in $a$), weakly separated in magnitude. The axes of $b$) have been rotated to facilitate a point of view that best highlights the billow.}
 \label{fig:Large_KH_sigma}
\end{sidewaysfigure}

The most dramatic K-H billows occur for the {\em {large}} ISW which consistently pulses such features towards its rear. Two such successive events, in the trailing edge of the {\em {large}} ISW, are shown in the instantaneous two-dimensional density contours on the $xz$-transect  in Fig. \ref{fig:Large_KH_sigma}a. The particular billows, fully overturning, extend to approximately $50$m in the vertical, which is significantly taller than the events shown in Fig. \ref{fig:Early_KH}. 

\par
\begin{sidewaysfigure}[h!]
 \centering
 \includegraphics[width=45pc]{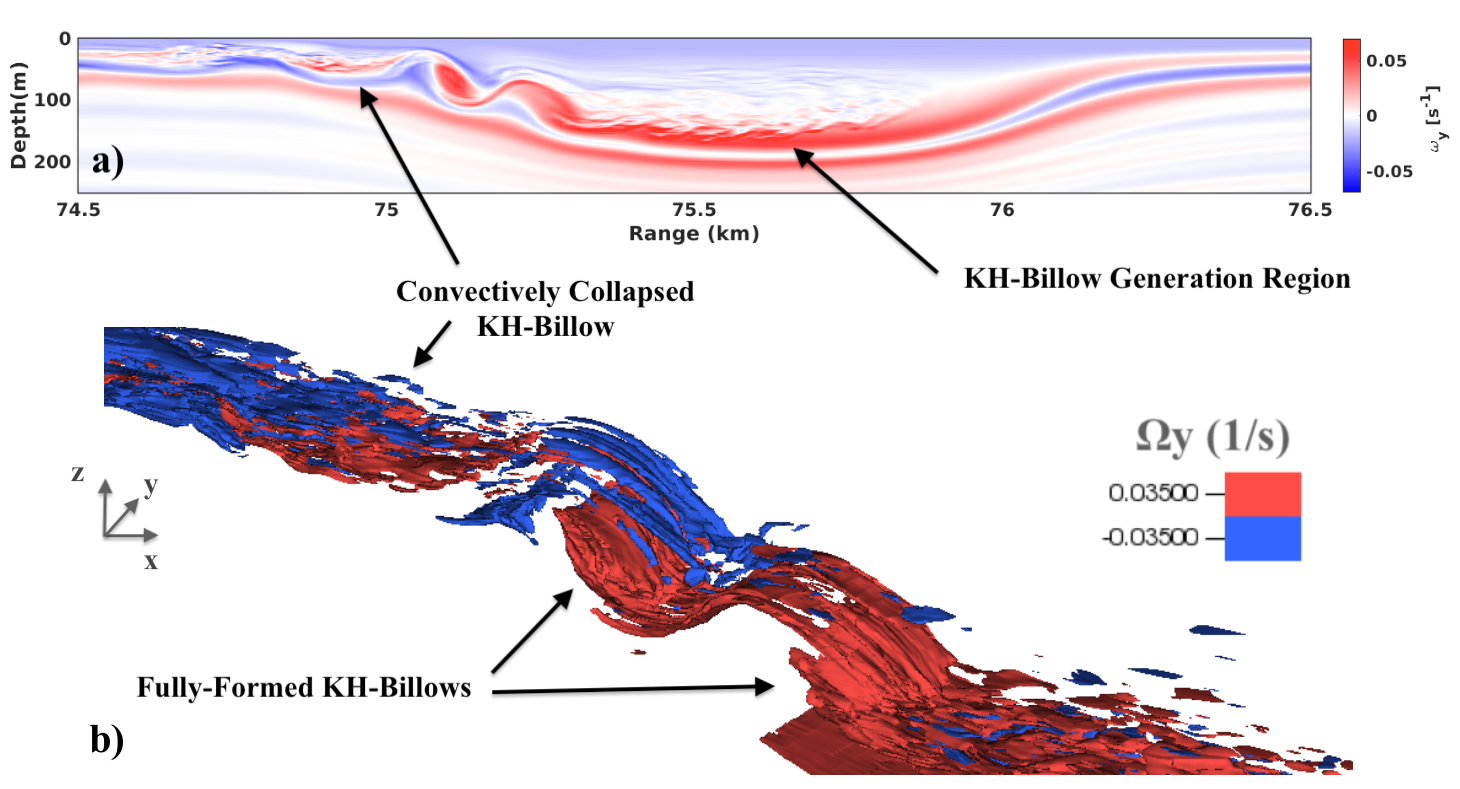}
 \caption{Transverse-averaged $y$-vorticity contours in the {\em{large}} ISW rear are shown in panel {\em{a}}. The presence of two distinct, robust vortical structures in the rear shoulder of the ISW indicates transverse coherence of the billows. Instantaneous three-dimensional snapshots of the large K-H billows' transverse vorticity field are also presented using two  $y$-vorticity isosurfaces of equal and opposite value (panel {\em{b}}). The ISW location is the same with that in Fig. \ref{fig:Large_KH_omega}.}
 \label{fig:Large_KH_omega}
\end{sidewaysfigure}

 An exploded view of the region outlined by the white box in \ref{fig:Large_KH_sigma}$a$, shows three-dimensional isosurfaces of density in Fig. \ref{fig:Large_KH_sigma}b with the two distinct roll-ups corresponding to the two K-H billows discussed above. Finer-scale lateral distortions of the isosurfaces are visible, representative of billow three-dimensionalization \citep{barad2010}, and are typically associated with 4-to-6 rib-like structures  along the billow periphery which are also visible in the transverse vorticity visualization of Fig. \ref{fig:Large_KH_omega}b. 

\par The apparent non-negligible transverse coherence of the two billows under consideration is further supported by visualizations of the corresponding spanwise-averaged $y$-vorticity field (Fig. \ref{fig:Large_KH_omega}a). The lateral structure of the $y$-vorticity field zoomed into these two billows is shown in Fig. \ref{fig:Large_KH_omega}b. While some variability in the form of rib-like structures is evident, the primary roll-up pattern remains clearly coherent. Note that the rib-like structures are associated with a vorticity component tangential to the billow periphery, i.e., streamwise and vertical vorticity, which is not plotted here, however, for the sake of visual clarity. The available resolution, combined with a spectral filter in the transverse direction (Appendix Aa and \citealt{diamessis2006}), are likely preventing further breakdown of the billow into turbulence at this stage of their evolution.

\par However, consistent with the absence of any overturning of the black isopycnal line on the rear shoulder of the wave or further behind it (Fig. \ref{fig:Large_KH_sigma}a), no other coherent patch of red vorticity lies ahead of these two billows along the shallowest, positive-signed, layer of $y$-vorticity in Fig. \ref{fig:Large_KH_omega}a. The three-layer vorticity structure is a direct imprint of the triple shear layer in the background current profile in the depth range of $[20,80]$m (Fig. \ref{fig:insitu_profs}a and b). At the rear ISW shoulder, one instead observes a finer-grained feature of red vorticity apparently linked to the breakdown of an earlier billow ascending from the wave trough.  A higher postprocessing file output rate (Appendix Ac) is needed to ascertain such a statement on billow evolution. Per the earlier observation on the additional resolution required for any transverse instabilities to transition into turbulence (e.g. see the work of \citealt{mashayek:11}), this particular billow breakdown is most likely due to its gravitational collapse \cite{smyth2000}.

\par Note that any billow-like features behind the wave, in the form of  $y$-vorticity roll-ups,  are actually observed in the middle, negative-signed, of the three $y$-vorticity layers. These counter-clockwise-rotating billows appear to be produced by the visibly steep roll-up of the negative $y$-vorticity layer by the clockwise-rotating billows generated at the wave trough prior to their breakdown (Fig. \ref{fig:Large_KH_omega}a). Finally, a weak negative y-vorticity pattern radiating towards the underlying weakly stratified layer is visible in the wave's wake; its investigation is deferred to future studies. 

\par
When compared to the in-situ observed billows of \cite{chang2021b}, the numerically simulated ones in Figs. \ref{fig:Large_KH_sigma} and \ref{fig:Large_KH_omega} are almost double in size. However, ISWs observed by Chang and coworkers were at least $25$m smaller in amplitude than our {\em {large}} simulated wave, which could partially explain this discrepancy in billow height. An additional factor behind this discrepancy may be the incomplete three-dimensional evolution of the simulated billows discussed above.

\par
In agreement with a number of past studies (\citealt{carr2011,barad2010,lamb2011,chang2021b}), K-H-like billows are found here to strictly form near the trough of the ISW. A subset of in-situ (\citealt{moum2003,Huang2022}) and laboratory (\citealt{carr2008b}) studies of similar, but not identical waves and configurations,  report billows originating not just at the trough but also the leading edge of the wave. The work of \citet{carr2008b} considered an ISW with a {\em {surface}}-attached recirculating core which would likely mix more efficiently the ISW's leading edge as contrasted to the subsurface core reported here which mixes the wave trough and rear. 

\section{Kinetic Energy Evolution} 
\label{sec:Kinetic_Energy}

\par
The Perturbation Kinetic Energy per unit mass of the fluctuating velocity components, integrated over a finite volume that encompasses the wave, is defined as

\begin{equation}
\mathbf{PKE}\: = \frac{\rho_0}{2}\int_{V} \: (u'^2 + v'^2 + w'^2)  \,dV~~.
\label{eq:KE}
\end{equation}

The quantities denoted in primes represent the fluctuation of each velocity component with respect to its y-averaged counterpart (which is denoted with $< y >$ hereafter) obtained from the instantaneous three-dimensional instantaneous field at a given location along the wave propagation track. Note that the y-averaged velocity field consists of the $z$-dependent and time-fixed background current and a $yz$-dependent component which is driven by the ISW itself and the larger-scale trace of any originally two-dimensional instabilities.

\par
Typically the quantity defined in Eq. (\ref{eq:KE}) would be regarded as “Turbulent Kinetic Energy” (\citealt{tennekes}) as it is primarily representative of the turbulent fluctuations in the flow. This is indeed the case for the associated fluctuating flow field that develops after the initial convective instability, as shown in Figs. \ref{fig:Baseline_Evolution_3D} and \ref{fig:Contours_All_Isws} per the complex finer-scale structure with appreciable scale separation found in the wave core. As discussed in Sec. \ref{sec:Shear_Inst}\ref{subsec:KH_Inst}, the K-H billows produced by shear instability do not become fully turbulent when evolving three-dimensionally, most likely due to insufficient lateral resolution and the use of a spectral filter. Hence, the term “Perturbation Kinetic Energy” (PKE) will be used hereafter. 

\par
\begin{figure}[H]
 \includegraphics[width=39pc]{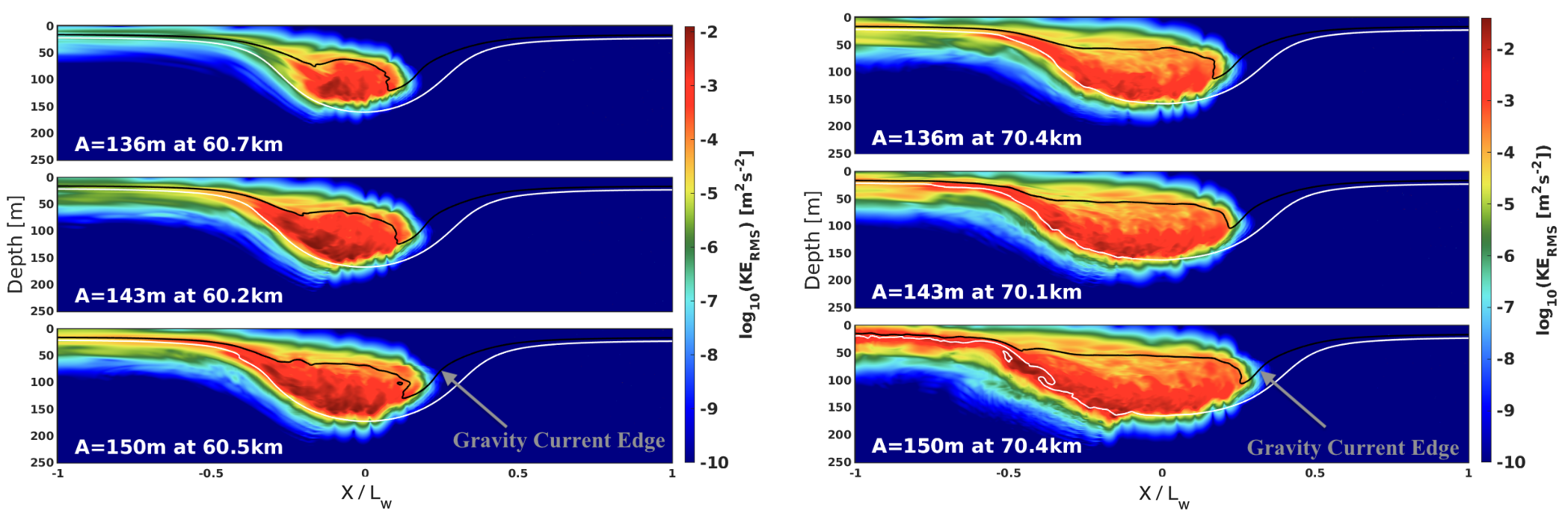}
  \caption{Contours of Perturbation Kinetic Energy, normalized by the constant $\rho_0$ factor (PKE/$\rho_0$), at $60$ (left) and $70$ km (right), centered relatively to the ISWs and presented as a function of distance from the centerline, normalized by the common approximate wavelength ($~1$km). Note the logarithmic scale on the color bars, that effectively reflect the associated maximum value computed in each location, which is not the same between the two locations along the propagation track. The PKE content of the more energetic fraction of each ISW interior is amplified for almost one order of magnitude on the plateau ($70km$) compared to the {\em {shallow}} mooring region ($60$km). The white and black contour lines correspond to those in red and grey in Figs. \ref{fig:Baseline_Evolution_3D} and \ref{fig:Contours_All_Isws} and are used to delineate the subsurface recirculating core and approximate edge of the gravity current and the ISW waveform, respectively. The full PKE evolution across all cases may be found in the supplementary animation.}
  \label{fig:PKE_RMS}
\end{figure}

Contours of the Root-Mean-Square transversed-average  PKE  for all three ISWs are shown using logarithmic color bars for two snapshots of wave evolution ($60$ and $70\:$km respectively) in Fig. \ref{fig:PKE_RMS}. Note that the logarithmic contours present the PKE values in  "m$^2$ $\cdot$ s$^{-2}$", as the constant \( \rho_0 \) from Eq. (\ref{eq:KE}) has been excluded. Although the larger the wave, the broader the along-wave extent of the high PKE region, all simulated waves show qualitatively similar behavior in their evolution (see also supplementary animations). At the {\em {shallow}} mooring ($60$km), the PKE is driven by velocity perturbations due to the convective instability, the associated entrainment of heavier water linked to the plunging isopycnals and the subsequent horizontally propagating gravity current. The PKE distribution is relatively uniform within at least the gravity current as delineated by the black-line isopycnal (that corresponds to the red isopycnal of Figs. \ref{fig:Baseline_Evolution_3D} and \ref{fig:Contours_All_Isws} with the color switch performed to enhance visibility) in Fig.  \ref{fig:PKE_RMS}. The top right (upper front) fraction of the ISW interior remains relatively inactive through the entire simulation, when compared to the mid-and-rear part of the wave core.

\par
The PKE values within the more energetic fraction of the interior of each ISW are amplified by almost one order of magnitude on the plateau ($70$km - right panels of Fig. \ref{fig:PKE_RMS}) as compared to the shallow mooring region ($60$km); the amplification is apparently due to the establishment of shear instabilities. Furthermore, larger PKE values are more localized near the trough and the tail of the wave, consistent with the trajectory of associated K-H billows (Figs. \ref{fig:Large_KH_sigma} and \ref{fig:Large_KH_omega}), particularly for the case of the {\em {large}} ISW.

\par
\begin{figure}[H]
 \includegraphics[width=39pc]{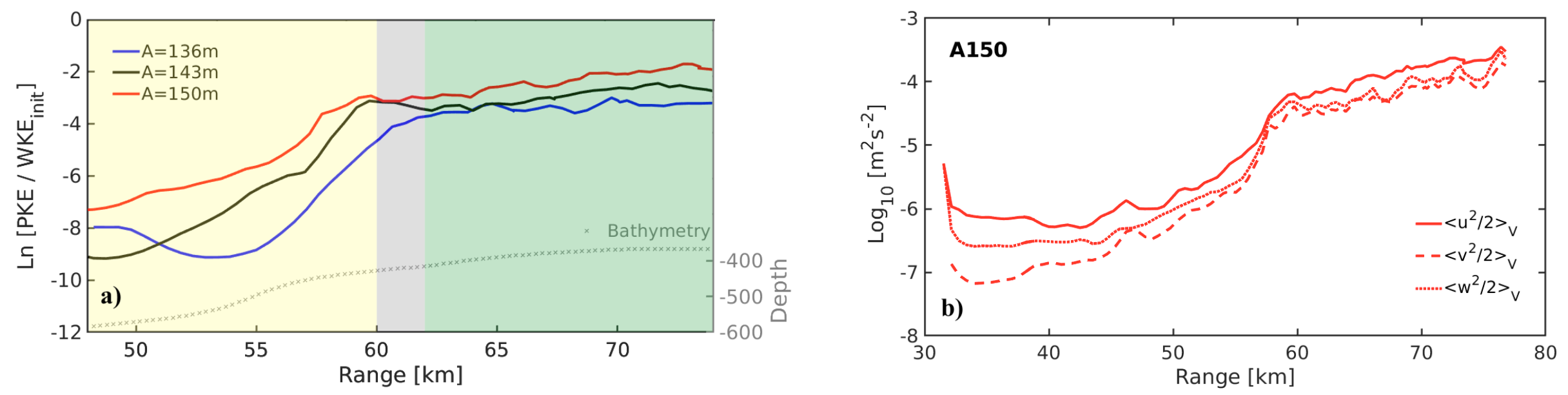}
 \caption{\textbf{a:}  Perturbation Kinetic Energy per unit mass integrated (Eq. (\ref{eq:KE})~) over a prescribed volume, $V$, horizontally centered around the ISW for all three ISWs, normalized by the kinetic energy of strictly the ISW (PKE is not accounted for) at time of three-dimensional simulation start-up as computed over the same volume, and \textbf{b:} PKE contribution of each fluctuating velocity component for the {\em {large}} wave. Both quantities are shown as a function of the ISW trough position along the transect. In panel a), the yellow-shaded area shows the convectively-unstable-dominated regime, while the green one denotes the shear-instability-dominated regime. The gray area represents the transition between the two regimes. The gray symbols on the bottom denote the bathymetry over this region, for reference. Note that panel a) uses the natural logarithm in $y-$axis to emphasize the exponential growth between [55, 60] km, while panel b) uses a base-10 logarithmic vertical axis to highlight the differences in the order of magnitude for each component. In panel b), each component is normalized with the volume of integration and constant density $\rho_0$, and is kept in dimensional form (m$^2$s$^{-2}$) to allow for a direct comparison with potential field measurements.}
 \label{fig:PKE_Evol}
\end{figure}

Figure \ref{fig:PKE_Evol}$a$ shows the evolution of the wave-integrated PKE (Eq. \ref{eq:KE}) per unit mas as a function of position along the propagation track. The values shown are normalized by the kinetic energy of the ISW immediately prior to the three-dimensional simulation initialization, which effectively does not include the contributions of the corresponding inserted three-dimensional perturbations and background current. The associated integration is performed within a two-wavelength-long ($\sim 2$km), $250$m-deep and $50$m-wide volume, which is horizontally centered at the ISW trough and encompasses the entire wave. Note that the particular integration volume does not account for the Kinetic Energy of the K-H-billow-driven wake (see Sec. \ref{sec:Wake}). A weak initial reduction in PKE amplitude evolution visible for the {\em{small}} wave, also observed for the {\em {baseline}} wave earlier in the transect (not shown here), is due to the transient adjustment of the inserted perturbations (Appendix Ab) to correlate with the preexisting ISW-dominated two-dimensional flow field and any reduction of the associated velocity magnitude because of the pressure projection step of the flow solver which ensures the perturbation velocity field is fully incompressible (Olivier Desjardins, pers. comm. and \citealt{diamantopoulos2022}). 

\par For all three ISWs, one subsequently discerns two regimes of PKE evolution along the transect; at any location, PKE magnitude increases with wave amplitude. In each regime, the wave-integrated PKE curve has a linear slope, indicating exponential growth since the vertical axis is plotted in natural logarithmic scale. Between $53$ and $60$km, steep, vigorous growth of the PKE is observed, apparently driven by the convective instability. The onset of this growth is delayed with decreasing ISW amplitude, consistent with the corresponding delay in instability onset (Fig. \ref{fig:UmaxC}). The slope of the PKE curve is comparable for all three waves in this first regime, particularly between $57$ and $60$km. Prior to that point,  a slower growth rate is observed for the {\em{large}} wave, presumably due to the residual perturbations from the weaker, shorter-lived convective instability event in deeper waters (Sec. \ref{sec:Conv_Evol}\ref{subsec:Conv_Br_All_ISWs}). Along the remaining part of the transect, including the plateau where shear instabilities dominate, the PKE continues to increase, albeit at a visibly slower rate. Recall that despite this reduced growth rate, the maximum PKE values are non-trivially larger than their deeper-water counterparts (left vs. right panels of Fig. \ref{fig:PKE_RMS}). However, in the latter case, PKE is confined near the pycnocline and does not permeate the bulk of the ISW core. In addition, for the {\em{large}} ISW, and particularly over the plateau, the PKE grows up to a point where it exceeds 15\% of the initial ISW-related kinetic energy. This suggests that the shear-instability-driven regime, which dominates this part of the transect, is highly energetic for this simulated wave.

\par
In Fig. \ref{fig:PKE_Evol}$b$, the contribution of each individual velocity component to PKE evolution is shown strictly for the {\em {large}} wave, noting the run's earlier time of transition to three-dimensional mode and perturbation insertion (Appendix Ab). The curves are plotted from this time onward. Note that for each component, the quantity shown is normalized with the constant $\rho_0$ of Eq. (\ref{eq:KE}) and averaged with the corresponding integration volume, to allow a potential direct comparison with future field measurements. Per the practice of inserting noise in only the $u$ and $w$ velocity components, the transverse ($v$) component does rapidly grow from a zero value and eventually becomes comparable in magnitude to the vertical ($w$) component. The three curves distinctly overlap only for a short propagation range ($\sim 55 - 60 \:$km) linked with the most intense phase of the convective instability development. On the remaining part of the transect, the along-wave ($u$) component is consistently elevated with respect to the other two components. Such behavior should be expected since the dominant contributor to vertical shear, and turbulence production \citep{pope}, in the along-wave propagation plane originates from the horizontal background and wave-induced current. Qualitatively similar behavior for the remaining two waves (not shown) is also observed. 

\section{Wake on the plateau}
\label{sec:Wake}
\par
High-amplitude ISWs have been observed to develop an actively turbulent wake throughout their trailing edge \citep{moum2003}. \cite{lien2012} observed an isopycnal salinity anomaly indicating a wake of mixed water, originating from the mixed ISW core. The particular study argued that this wake might be attributed to the leaking/detrainment of well-mixed fluid from the recirculating turbulent core of the convectively unstable ISW. More recently, \cite{lucas2022} examined wakes in ISWs of elevation proposed to be produced by transverse instabilities in the wave interior.

\par
\begin{figure}[H]
 \centering
 \includegraphics[width=39pc]{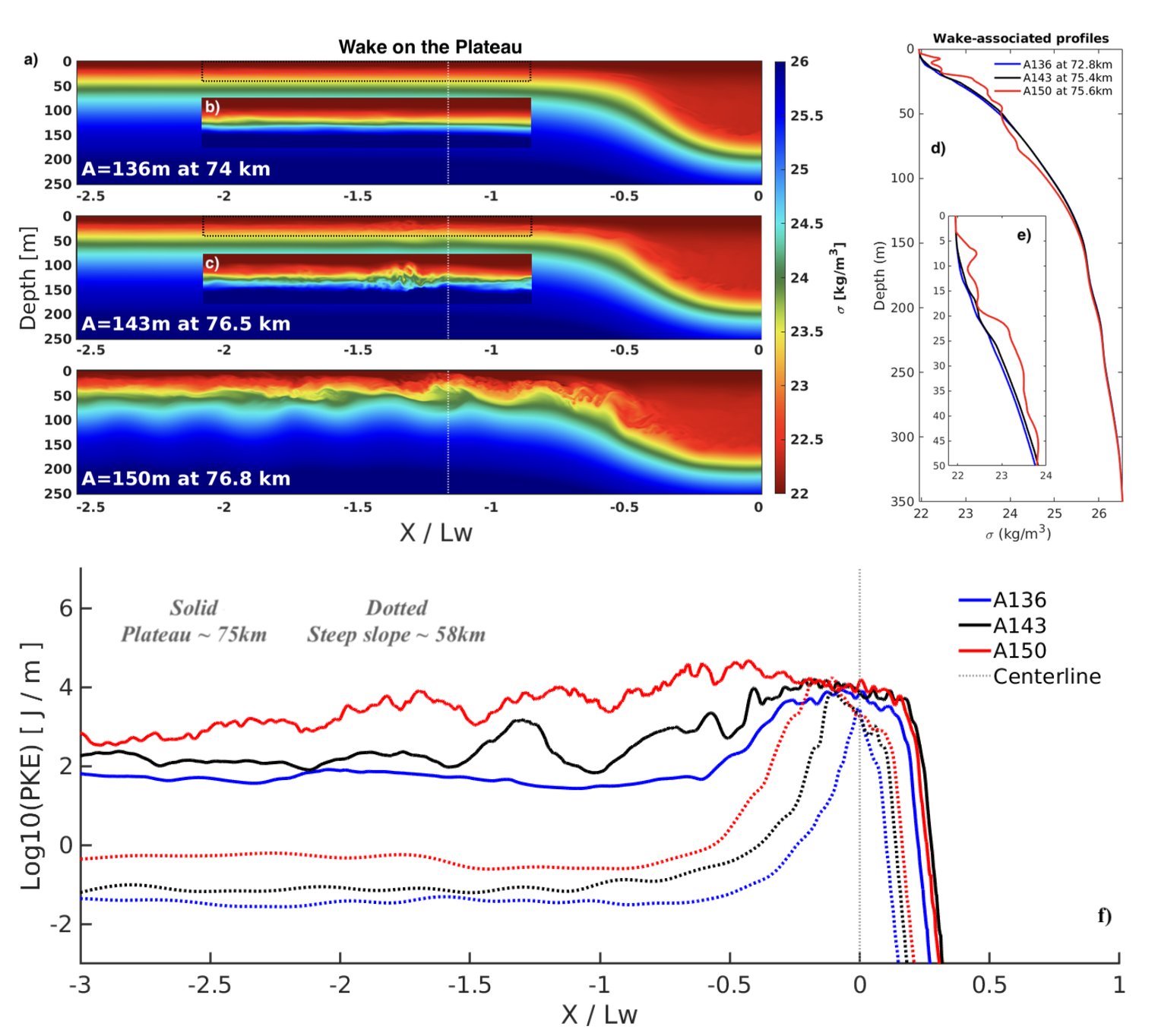}
\caption{Midplane density contours within the wake produced on the plateau for all three simulated waves (\textbf{a}). Panels (\textbf{b}) and (\textbf{c}) visualize the corresponding black-outlined boxes in (\textbf{a}) with a reduced range in the colorbar limits ([22 23]kg/m$^3$). The vertical white dotted lines indicate the location of profiles sampled and shown in (\textbf{d}). A zoomed view of the near-surface region (top 50m) is shown in \textbf{e}, emphasizing the large density disturbances for the {\em{large}} wave. Part (f) shows the surface $YZ-$ integral of the PKE of the wake as a function of horizontal distance relative to the centerline (vertical gray dotted line). Solid lines correspond to ISW positions along the plateau (that map to those on panel a), while dotted lines are computed near 58km of the transect where the convective instability dominates.}
 \label{fig:Wake}
\end{figure}

The wakes that develop over the plateau for all simulated ISWs are shown in Fig. \ref{fig:Wake}a, by visualizing density contours on the stream-depth mid-plane. The field-of-view consists of a $2.5 \: km$-long window which extends behind each wave's trough, shown via the normalized distance from the centerline (the nominal wavelength $L_x=1$km is used; see Tab. 1). The actual centerline location of each wave along the transect is shown at the bottom left corner of each panel. An initial inspection of Fig. \ref{fig:Wake}a) suggests that a distinct wake forms only behind the {\em {large}} wave. However, the presence of a less active, and more vertically confined, wake is identifiable for both  {\em {small}} and {\em {baseline}} ISWs when the color-bar limits are adjusted as shown in the inset Figs. \ref{fig:Wake}b and \ref{fig:Wake}c respectively, which correspond to the black-outlined box in the larger field-of-view.

\par
Unlike \cite{lien2012} who suggest that the wake results from convective instability of the ISW, Figs. \ref{fig:Large_KH_sigma}, \ref{fig:Large_KH_omega} \& \ref{fig:Wake}a indicate a distinct link of the simulated wake to the shear instabilities and clockwise-rotating K-H billows generated at the wake trough. The {\em {large}} wave (sampled here at a later time as compared to Fig. \ref{fig:Large_KH_sigma}a) in particular develops a uniquely visible wake in its rear, consisting of interfacial wave-like features whose wavelength is consistent with the spacing of the clockwise-rotating billows generated at the ISW trough (Fig. \ref{fig:Large_KH_sigma}a and c). Note, however, that any density overturns \textit {in the wake} of the \textit{large} ISW indicate stirring by a counter-clockwise rotation as they are caused by the subsequenet evolution of negative $y$-vorticity features produced at the ISW rear shoulder (Fig. \ref{fig:Large_KH_omega}a). Although these counter-clockwise billows are clearly left behind the ISW, determining whether they propagate away from the wave and at what speed would require a higher postprocessing file output rate than the one currently used (see Appendix Ac). 

\par The vertical overturns and turbulent motions linked to billow-breaking suggest additional active mixing of the background stratified water column for long distance at the wake of the ISW ($\mathcal{O}(10$km)). Finally, for the case of only the {\em{large}} wave, large-scale overturns on the $y-z$ plane, up to $15$m-tall, are observed in the ISW wake (not shown).

\par
Instantaneous density profiles sampled in each wave's {\em {wake} } ($1.14$km behind the trough, also shown by the vertical white dotted lines in Fig. \ref{fig:Wake}$a$) are shown in Fig. \ref{fig:Wake}$d$. The location of the wake-profile sampling is chosen appropriately aiming to juxtapose the large density inversions produced by the {\em {large}} wave, especially as compared to the {\em {small}} and {\em {baseline}} simulated waves. The earlier along-transect sampling location for the {\em {small}} ISW (as shown in bottom-left corner of panel a) is because the particular simulation was stopped at a slightly earlier time (its last output is actually shown in Fig. \ref{fig:Wake}) with respect to the {\em {large}} and {\em {baseline}} waves.
\par
The profiles show a highly perturbed water column only for the {\em {large}} wave (Fig. \ref{fig:Wake}$d$). The associated deviations from the reference background profile for the {{\em small}} and {\em {baseline}} ISW are negligible, as these profiles remain almost unperturbed with respect to their initial deep-water structure. However, for the {\em {large}} wave, a density inversion larger than $1\:$kg/m$^3$ between $20$ and $30$m of depth is observed. Such high deviations are potentially linked with high available potential energy and continuous turbulent mixing of the water column which would then non-trivially modify the background state of potential energy \citep{winters1995}.

\par  
Figure \ref{fig:Wake}f shows the $yz$-surface integral of the perturbation kinetic energy (PKE) as a function of horizontal distance from the respective wave centerline (denoted by the gray vertical dotted line)  within each ISW wake on the plateau (solid lines). Integration at each $x$-location is performed over a stream-span surface spanning the entire computational domain width and corresponding water column depth. The \(x\)-axis in Fig. \ref{fig:Wake}f extends over a length of $4km$ length, as compared to the $2.5km$ horizontal axis of the \(\sigma\) contour-plot in panel (a) of the same figure. 

\par 
An important consideration when examining Fig. \ref{fig:Wake}f, namely per the lateral coherence of the K-H billows in Fig. \ref{fig:Large_KH_omega}, is that the y-averaged signal, and effectively the associated larger-scale kinetic energy, of each billow have been subtracted from the PKE calculation. Therefore, the computed PKE represents only the finer-scale structures associated with billow breaking.

\par  
Although the PKE of all simulated ISWs on the plateau nearly overlaps near the wave’s centerline, significant differences emerge in the wake (solid lines in panel f), particularly between the {\em{small} and {\em{large} ISWs, up to three orders of magnitude. The PKE curves for the {\em{large} and {\em{baseline} waves exhibit a distinctly oscillatory behavior in the interval $X/L_W \in [-2,-0.5]$. These oscillations appear to be convectively breaking KH-billows shed by the ISW or generated in ISW-wake itself in the negative-sign-vorticity shear layer of the background profile (see the discussion in Sec. \ref{sec:Shear_Inst} \ref{subsec:KH_Inst}). In contrast, no such oscillations are observed for the {\em{small}} wave which supports weak billows or billows that have been found to break convectively before being shed in the rear of ISW waveform.

\par  
For comparison, the corresponding distribution of PKE for each wave at $58$km along the transect, when convective instability dominates (see also Fig. \ref{fig:PKE_Evol}a) but KH-billows have not yet emerged, is also shown with dotted lines in Fig. \ref{fig:Wake}f. As suggested by the contour plots in the left panel of Fig. \ref{fig:PKE_RMS}, the $yz$-integrated PKE signal for the ISWs at $58$km is dominated by convectively-generated PKE originating from the rear shoulder of the wave and extending past its centerline to the front shoulder, with minimal contribution from the wave wake (three-to-four orders of magnitude smaller). 
\par  
Notably, at $58$km, the PKE inside the wave core is at best (for the {\em{large}} wave) comparable to the shear-instability-driven PKE values in the wave rear on the plateau at $75$km. The PKE in the actual ISW wake, at $75$km, is notably larger than the corresponding values in the ISW core at $58$km, over a distance behind the wave which increases with wave amplitude. This substantial amount of PKE in the shear-instability-driven motions within the ISW wake further suggest (along with the associated density overturns shown in Fig. \ref{fig:Wake}f) that these motions are the primary mechanism responsible for any long-term modification of the background water column.

\par Finally, note that the investigation of the evolution of theshear-instability-generated wake, over even longer distances behind the wave with respect to those examined here, is constrained by the use of the overlapping window approached discussed in Sec. \ref{sec:Method}\ref{subsec:Sim_descr}\ref{subsubsec:Comp_domain} and Fig. \ref{fig:domain}. This constraint is the most evident for the large ISW examined in this study. To this end, longer overlapping windows, entailing higher computational cost, would be needed and would be the topic of future research.

\section{Discussion and Concluding Remarks}
\label{sec:Discussion}

\par
In this paper, the shoaling of high-amplitude internal solitary waves (ISWs) over a select, near-isobath-normal, transect of the South China Sea has been examined using turbulence-resolving simulations (at spatial and temporal resolutions of $[0.21,1]$m,  across all three spatial directions, and $0.2$s, respectively) equipped with realistic profiles of background current and stratification. The use of a state-of-the-art numerical modeling approach \citep{diamantopoulos2022} has enabled the reproduction, at such high-resolution over the entire wave propagation track (extending over $80$km and spanning depths from $921$m to $365$m), of key features of the corresponding in-situ observed phenomenology (\citealt{lien2012,lien2014} $\&$ \citealt{chang2021a,chang2021b}). For three simulated ISWs, of varying initial amplitude (corresponding to maximum isopycnal displacements of $136, 143$, and $150$m), the evolution of the wave-scale response has been captured in detail along with the development of associated convective and shear instabilities and the larger turbulent scales therein. 

\par 
The convective instability and the resulting subsurface recirculating core formation occur in a similar fashion across all three waves. However, the onset of convective instability occurs earlier along the transect with increasing initial wave amplitude. The enhanced spatial resolution of the simulations reveals a particular aspect of the convective instability evolution and the corresponding finer-scale lateral structures, previously not reported in the field: for all simulated waves, the convective instability leads to the entrainment of heavier water from the rear shoulder of the wave, which plunges into the wave interior. Upon sufficient accumulation of heavier water inside the rear of the wave, a gravity current is then initiated which advances to the ISW's leading edge, further advecting the heavier water with as entrainment in the rear of the wave continues. The gravity current's propagation speed is nearly constant across all three simulated waves. The waveform remains distinctly symmetric throughout this process. Both the convective plunging and gravity current generate turbulence across the entire wave interior which visibly mixes the rear of the ISW but also weakens the stratification at the wave trough. 

\par At this stage of ISW shoaling, local patches of Richardson number below $0.1$  emerge, giving rise to shear instability and Kelvin-Helmholtz (K-H) billows. Although the amplitude of the three simulated waves varies only by an additive factor of $7$m, K-H billow size and their signature behind the wave grow nonlinearly with wave amplitude. $50$m-tall billows are observed within the largest simulated wave upon its arrival on the shallower water plateau at 360m depth. A visible and highly energetic wake, persisting 5km behind the wave, is observed in this case as contrasted to the two lower-amplitude ISWs where only weak intermittent bursts of turbulent-like motion occur in the lee of the wave. From a perturbation kinetic energy (PKE) perspective, the wake appears to be directly linked to the highly energetic K-H-billows and their finer-scale three-dimensional structures that emerge from them; the larger-scale roll-ups continue to overall maintain their transverse coherence. At distances up to one ISW wavelength behind the wave, the PKE within the wake can become as high as 15\% of the kinetic energy of the ISW-induced velocity at the time of onset of convective instability in deeper waters. 

\par The perturbation kinetic energy of the shear-instability-dominated stage is focused along the billow propagation path, i.e., the pycnocline and wake but is at least one order of magnitude larger than that produced by its wave-filling counterpart through the antecedent convective instability. Each of the two instabilities imparts its own near-exponential growth rate in the wave-integrated along-shoaling-track evolution of the perturbation kinetic energy further marking the delineation between two separate evolution regimes in the shoaling ISWs. 

\par Recently-deployed towed chains of conductivity-temperature-density (CTD) sensors, used to capture shear instabilities in topographic lee-waves within the Kuroshio \citep{yeh:24}, may be the optimal measurement tool to capture in the field the phenomena reported here. \citet{vladoiu:22} suggest that the uncertainty of such an individual CTD sensor is $0.003~kg/m^3$ which is sufficient to resolve the density differences associated with the convective-and-shear-instability-driven phenomena reported here. The main challenges a deployment with a towed CTD chain would be faced with in capturing in two dimensions the structure of the features of interest are mainly linked to the maximum attainable resolution in the vertical and along-track directions and the ability to avoid along-track spatial versus temporal aliasing, i.e., whether any measured variation is actually along-track or temporal in its origin.

\par
The results reported here serve as the springboard for ongoing and future analysis of the existing datasets and future larger-scale high-accuracy/resolution simulations to investigate a number of open questions that have emerged in this manuscript in regard to the turbulence dynamics of shoaling ISWs on the continental slope. Examples of such questions, linked to limitations of computational resources facing the large-scale parallel turbulence-resolving simulations reported here, are determining the role of lateral domain dimension on convective instability development and exploring the development of the shear-instability-driven wake over longer distances behind the ISW. Once these and other remaining open questions are addressed, the ground should be mature for modeling of relevant, and more complex and diverse, ISW-driven processes in shallower waters as recently reported through several recent studies \citep{colosi2018,Davis2020,Ramp2022a,Ramp2022b,Sinnett2022,Whitwell2024}.

\clearpage
\acknowledgments
The authors are grateful to Professors Marek Stastna, Jim Riley, Steve de Bruyn Kops, Greg Ivey, Larry Armi, and Olivier Desjardins for several insightful conversations on turbulence in breaking internal solitary waves. This work has been primarily supported by the National Science Foundation (NSF) grant OCE-1634257. Additional support by grants NSF-OCE 1948251 and NSF-OCE 1634182 is also acknowledged. This work used Stampede2 at TACC through allocation TG-EES200010 from the Extreme Science and Engineering Discovery Environment (XSEDE), which was supported by National Science Foundation grant number 1548562. Moreover, this work used Stampede2 at TACC through allocation TG-EES200010 from the Advanced Cyberinfrastructure Coordination Ecosystem: Services $\&$ Support (ACCESS) program, which is supported by National Science Foundation grants 2138259, 2138286, 2138307, 2137603, and 2138296.

%
%
\datastatement Data used to produce the main analysis and figures contained in this manuscript are available by request. Further details can be found at \url{https://diamessis.github.io}.





\appendix[A] 
\label{adx:NumTool}


\appendixtitle{Numerical Tool}

\subsection{Numerical Stability}

\par
Numerical stability of the SEM solver is enabled explicitly through the application of a $10^{th}$ and $15^{th}$ order exponential spectral filter in the non-homogeneous x-z plane and periodic y-direction respectively (\citealt{diamessis2005} and \citealt{diamantopoulos2021}), equivalent to the use of hyperviscous operators (\citealt{Boyd1998,gottlieb2001,winters:15}). Each velocity component and density perturbation are filtered once every time-step after the computation of the non-linear and advective terms respectively \citep{diamantopoulos2022}, in contrast to \cite{riverarosario2022}. Additional stabilization is achieved in the $x-z$ directions via ``polynomial dealiasing'' \citep{kirby2003}, of the non-linear and advective terms (\citealt{malm2013,diamantopoulos2022}). Lastly, an artificial Rayleigh-type damper, two ISW-width thick, is applied to the left and right boundaries to eliminate any possible reflection from the incoming ISW \citep{riverarosario2020}. 

\subsection{Noise Insertion}
\par
Specifically, white Gaussian noise with a standard deviation $\sigma \: = \: 10^{-2}$ and maximum amplitude $A_{noise} = 0.015$ (units are m/s and kg/m$^3$ for velocity and density, respectively), is inserted immediately after the transition to three-dimensional mode. The inserted perturbations cover the entire $xz$-plane in the computational domain, with the exception of thin zones along the top, bottom and lateral boundaries of the domain as defined by a masking function (aimed to prevent numerical instabilities) \citep{diamantopoulos2021}, for select low-wavenumber transverse Fourier modes. In particular, the spectral support of this perturbation field  in $y$-Fourier space is restricted over the first seven non-zero modes of only $u$, $w$, and $\rho'$ components. The energy content of any higher transverse Fourier modes builds up naturally through nonlinear interactions and the resulting downscale energy cascade. The initial $v$ perturbation field is created through continuity, and the IEEB solver's projection step \citep{diamantopoulos2022}, similar to \cite{fringer2003}. 

\subsection{Computational resources} 
The massively parallel three-dimensional shoaling ISW turbulence-resolving simulations were conducted on the NSF ACCESS’s allocated HPC systems (\citealt{XSEDE} and \citealt{ACCESS}). For each run, the parallel execution of the flow solver on each computational window used 6144 cores evenly across 128 nodes on Stampede2 at the Texas Advanced Computing Center (TACC). The total number of timesteps for each simulation depends on the three-dimensional initialization location along the transect of each ISW, while the wall-clock time per timestep depends on the number of iterations of the direct/iterative solver for the pressure (\citealt{diamantopoulos2022}). On average, the total number of timesteps after the switch to three-dimensional mode  was $120000$, the wall-clock time per timestep $8$s, and the total simulation wall-clock time was $11$ days. Details of the outputting frequency are given in Appendix Ad.

\subsection{3D Outputs}
For all three simulations, the analysis reported in Sec. 3-6 is based on explicitly postprocessing large three-dimensional simulation output files. This large file size (24 Gb per file) imposes a practical constraint on the frequency at which these files are outputted during the simulation. As such, files are outputted every $[1224,2913]$ time steps (depending on the adaptively set time step value; see Sec. \ref{sec:Method}\ref{subsec:Sim_descr}\ref{subsubsec:Comp_domain}) which corresponds to an output rate in the range of $[270,339]s$ (a factor of $[0.5,0.65]$ of ISW time scale) in simulation time units and an along-propagation-track spacing of $[367,607]~m$ (a factor of $[0.35, 0.6]$ of ISW wavelength). This relatively coarse output rate limits the temporal resolution with which one can examine, in a postprocessing sense, the evolution of the K-H billows generated at the ISW-trough (Sec. \ref{sec:Shear_Inst}\ref{subsec:KH_Inst}).

\appendix[B] 
\label{adx:2Dvs3D}

\appendixtitle{Simulation comparison in two and three dimensions}

So far, in the published literature, any computational studies of the shoaling of ISWs over gentle slopes and long domains have largely relied on two-dimensional simulations \citep{lamb2015,riverarosario2020}. Although no results are explicitly shown, here, we summarize the main similarities and differences between two-dimensional simulations and their turbulence-resolving counterparts in three-dimensions (initiated from the onset of convective instability), as restricted to the {\em {baseline}} wave ($A = 143 $m) per the comparison already offered in \citet{diamantopoulos2021} and \citet{riverarosario2022}. 

\par
Per the calculation approaches outlined in Sec. \ref{sec:Conv_Evol}\ref{subsec:Conv_Br_All_ISWs}, in terms of the evolution of the maximum wave-induced horizontal velocity, $u_{max}$ relative to the ISW propagation speed, $C$, the onset of the convective instability as denoted by $u_{max} > C$ (see Sec. \ref{sec:intro}), occurs at almost the same location for the two and three-dimensional runs, as emphasized in \citet{diamantopoulos2021}. Satisfaction of the particular instability criterion is negligibly affected by any early-time finer-scale turbulent motions. In a similar vein, a comparison (not shown in this paper) of the amplitude $A$ (defined as the maximum ISW-induced vertical isopycnal displacement $\eta_{max}$) and wavelength $L_w$, calculated according to \citet{koop1981} and \citet{riverarosario2022} as,
 \begin{equation}
 \label{eq:lw}
 L_w = \frac{1}{A} \int_{-\infty}^{\infty} \eta_{max}(x) \,dx ~,
 \end{equation}
for the two-versus-three-dimensional runs reveals nearly identical behavior as a function of the downstream propagation, corroborating the findings of the lower-resolution simulations of \cite{riverarosario2022}. In conclusion, any three-dimensional effect does not affect dramatically the primary length and velocity scales of the ISW over the entire evolution.

\par
However, as demonstrated in \citet{diamantopoulos2021}, a focused examination of the wave core structure reveals differences in the evolution of the structure of convective instability between simulations in two and three dimensions. The wave interior (also referred to as the wave core) is also significantly more mixed, largely due to the larger-scale turbulence captured by the three-dimensional runs (Figs. \ref{fig:Contours_All_Isws} and \ref{fig:PKE_RMS}). Most importantly, to this end, shear instability occurs only in turbulence-resolving simulations. 

\par Note that a separate two-dimensional run for the baseline wave was performed with the perturbation insertion approach outlined in Sec. \ref{sec:Method}\ref{subsec:Sim_descr}\ref{subsubsec:initialize}. A rapid decay of the perturbations took place with no impact on the longer-development of the ISW core structure.







\bibliographystyle{ametsocV6}
\bibliography{references,pd-gs}

\begin{thebibliography}{105}
\providecommand{\natexlab}[1]{#1}
\providecommand{\url}[1]{\texttt{#1}}
\renewcommand{\UrlFont}{\rmfamily}
\providecommand{\urlprefix}{URL }
\expandafter\ifx\csname urlstyle\endcsname\relax
  \providecommand{\doi}[1]{https://doi.org/\discretionary{}{}{}#1}\else
  \providecommand{\doi}{https://doi.org/\discretionary{}{}{}\begingroup
  \urlstyle{rm}\Url}\fi
\providecommand{\eprint}[2][]{\url{#2}}

\bibitem[{Aigner et~al.(1999)Aigner, Broutman,, and Grimshaw}]{aigner1999}
Aigner, A., D.~Broutman, and R.~Grimshaw, 1999: Numerical simulations of
  internal solitary waves with vortex cores. \textit{Fluid Dyn. Res.},
  \textbf{25~(6)}, 315, \doi{10.1016/S0169-5983(98)00046-X}.

\bibitem[{Barad and Fringer(2010)Barad, and Fringer}]{barad2010}
Barad, M., and O.~Fringer, 2010: Simulations of shear instabilities in
  interfacial gravity waves. \textit{J.\ Fluid\ Mech.}, \textbf{644}, 61 -- 95,
  \doi{10.1017/S0022112009992035}.

\bibitem[{Boerner et~al.(2023)Boerner, Deems, Furlani, Knuth,, and
  Towns}]{ACCESS}
Boerner, T.~J., S.~Deems, T.~R. Furlani, S.~L. Knuth, and J.~Towns, 2023:
  Access: Advancing innovation: {NSF}’s advanced cyberinfrastructure
  coordination ecosystem: Services \& support. \textit{Practice and Experience
  in Advanced Research Computing}, Association for Computing Machinery, New
  York, NY, USA, 173–176, PEARC '23, \doi{10.1145/3569951.3597559}.

\bibitem[{Boyd(1998)}]{Boyd1998}
Boyd, J.~P., 1998: Two comments on filtering (artificial viscosity) for
  {C}hebyshev and {L}egendre spectral and spectral element methods: Preserving
  boundary conditions and interpretation of the filter as a diffusion.
  \textit{J.\ Comput.\ Phys.}, \textbf{143~(1)}, 283--288,
  \doi{https://doi.org/10.1006/jcph.1998.5961}.

\bibitem[{Canuto et~al.(2007)Canuto, Hussaini, Quarteroni,, and
  Zang}]{canuto2007spectral}
Canuto, C., M.~Hussaini, A.~Quarteroni, and T.~Zang, 2007: \textit{Spectral
  Methods: Evolution to Complex Geometries and Applications to Fluid Dynamics}.
  Scientific Computation, Springer Berlin Heidelberg,
  \doi{10.1007/978-3-540-30728-0}.

\bibitem[{Carpenter et~al.(2007)Carpenter, Lawrence,, and
  Smyth}]{carpenter2007}
Carpenter, J., G.~Lawrence, and W.~Smyth, 2007: Evolution and mixing of
  asymmetric {H}olmboe instabilities. \textit{J.\ Fluid\ Mech.}, \textbf{582},
  103 -- 132, \doi{10.1017/S0022112007005988}.

\bibitem[{Carr et~al.(2017)Carr, Franklin, King, Davies, Grue,, and
  Dritschel}]{carr2017}
Carr, M., J.~Franklin, S.~E. King, P.~A. Davies, J.~Grue, and D.~G. Dritschel,
  2017: The characteristics of billows generated by internal solitary waves.
  \textit{J.\ Fluid\ Mech.}, \textbf{812}, 541–577,
  \doi{10.1017/jfm.2016.823}.

\bibitem[{Carr et~al.(2008)Carr, Fructus, Grue, Jensen, Davies,, and
  Dalziel}]{carr2008b}
Carr, M., D.~Fructus, J.~Grue, A.~Jensen, P.~Davies, and S.~Dalziel, 2008:
  Convectively induced shear instability in large amplitude internal solitary
  waves. \textit{Phys.\ Fluids}, \textbf{20}, \doi{10.1063/1.3030947}.

\bibitem[{Carr et~al.(2012)Carr, King,, and Dritschel}]{carr2012}
Carr, M., S.~King, and D.~Dritschel, 2012: Instability in internal solitary
  waves with trapped cores. \textit{Phys.\ Fluids}, \textbf{24},
  \doi{10.1063/1.3673612}.

\bibitem[{Carr et~al.(2011)Carr, King,, and Dritschel}]{carr2011}
Carr, M., S.~E. King, and D.~G. Dritschel, 2011: Numerical simulation of
  shear-induced instabilities in internal solitary waves. \textit{J.\ Fluid\
  Mech.}, \textbf{683}, 263–288, \doi{10.1017/jfm.2011.261}.

\bibitem[{Carter et~al.(2005)Carter, Gregg,, and Lien}]{carter2005}
Carter, G.~S., M.~C. Gregg, and R.-C. Lien, 2005: Internal waves, solitary-like
  waves, and mixing on the {M}onterey {B}ay shelf. \textit{Cont. Shelf Res.},
  \textbf{25~(12)}, 1499--1520, \doi{10.1016/j.csr.2005.04.011}.

\bibitem[{Chang(2021)}]{chang2021c}
Chang, M.-H., 2021: Marginal instability within internal solitary waves.
  \textit{Geophys. Res. Lett.}, \textbf{48~(9)}, e2021GL092\,616,
  \doi{10.1029/2021GL092616}.

\bibitem[{Chang et~al.(2021{\natexlab{a}})Chang, Lien, Lamb,, and
  Diamessis}]{chang2021a}
Chang, M.-H., R.-C. Lien, K.~Lamb, and P.~Diamessis, 2021{\natexlab{a}}:
  Long‐term observations of shoaling internal solitary waves in the
  {N}orthern {S}outh {C}hina {S}ea. \textit{J. Geophys. Res.: Oceans},
  \textbf{126}, \doi{10.1029/2020JC017129}.

\bibitem[{Chang et~al.(2011)Chang, Lien, Yang,, and Tang}]{chang2011}
Chang, M.-H., R.-C. Lien, Y.~Yang, and T.~Tang, 2011: Nonlinear internal wave
  properties estimated with moored adcp measurements. \textit{J. Atmos. Oceanic
  Technol. - J ATMOS OCEAN TECHNOL}, \textbf{28}, 802--815,
  \doi{10.1175/2010JTECHO814.1}.

\bibitem[{Chang et~al.(2021{\natexlab{b}})}]{chang2021b}
Chang, M.-H., and Coauthors, 2021{\natexlab{b}}: Direct measurements reveal
  instabilities and turbulence within large amplitude internal solitary waves
  beneath the ocean. \textit{Commun. Earth Environ.}, \textbf{2~(1)}, 15,
  \doi{10.1038/s43247-020-00083-6}.

\bibitem[{Cheng et~al.(2024)Cheng, Chang, Yang, Jan, Ramp, Davis,, and
  Reeder}]{cheng2024}
Cheng, Y.-H., M.-H. Chang, Y.~J. Yang, S.~Jan, S.~R. Ramp, K.~A. Davis, and
  D.~B. Reeder, 2024: Insights into internal solitary waves east of {Dongsha
  Atoll} from integrating geostationary satellite and mooring observations.
  \textit{J.\ Geophys.\ Res.}, \textbf{129~(8)}, \doi{10.1029/2024JC021109}.

\bibitem[{Colosi et~al.(2018)Colosi, Kumar, Suanda, Freismuth,, and
  MacMahan}]{colosi2018}
Colosi, J.~A., N.~Kumar, S.~H. Suanda, T.~M. Freismuth, and J.~H. MacMahan,
  2018: Statistics of internal tide bores and internal solitary waves observed
  on the inner continental shelf off {P}oint {S}al, {C}alifornia. \textit{J.
  Phys. Oceanogr.}, \textbf{48~(1)}, 123 -- 143, \doi{10.1175/JPO-D-17-0045.1}.

\bibitem[{Cortes et~al.(2014)Cortes, Rueda,, and Wells}]{cortes:14}
Cortes, A., F.~J. Rueda, and M.~G. Wells, 2014: Experimental observations of
  the splitting of a gravity current at a density step in a stratified water
  body. \textit{J. Geophys. Res.}, \textbf{119~(2)}, 1038--1053,
  \doi{10.1002/2013JC009304}.

\bibitem[{Davis et~al.(2020)Davis, Arthur, Reid, Rogers, Fringer, DeCarlo,, and
  Cohen}]{Davis2020}
Davis, K.~A., R.~S. Arthur, E.~C. Reid, J.~S. Rogers, O.~B. Fringer, T.~M.
  DeCarlo, and A.~L. Cohen, 2020: Fate of internal waves on a shallow shelf.
  \textit{J. Geophys. Res.: Oceans}, \textbf{125~(5)},
  \doi{10.1029/2019JC015377}.

\bibitem[{Diamantopoulos(2021)}]{diamantopoulos2021}
Diamantopoulos, T., 2021: A high-order hybrid flow solver for the simulation of
  non-linear internal waves in long complex domains: exploring the turbulent
  aspects of a recirculating core in a shoaling internal solitary wave of
  depression. \doi{10.7298/j0ff-3h90}.

\bibitem[{Diamantopoulos et~al.(2022)Diamantopoulos, Joshi, Thomsen,
  Rivera-Rosario, Diamessis,, and Rowe}]{diamantopoulos2022}
Diamantopoulos, T., S.~Joshi, G.~Thomsen, G.~Rivera-Rosario, P.~Diamessis, and
  K.~Rowe, 2022: A high accuracy/resolution spectral
  element/{F}ourier-{G}alerkin method for the simulation of shoaling non-linear
  internal waves and turbulence in long domains with variable bathymetry.
  \textit{Ocean Model}, \textbf{176}, 102\,065,
  \doi{10.1016/j.ocemod.2022.102065}.

\bibitem[{Diamessis et~al.(2005)Diamessis, Domaradzki,, and
  Hesthaven}]{diamessis2005}
Diamessis, P., J.~Domaradzki, and J.~Hesthaven, 2005: A spectral multidomain
  penalty method model for the simulation of high {R}eynolds number localized
  stratified turbulence. \textit{J.\ Comput.\ Phys.}, \textbf{202}, 298--322,
  \doi{10.1016/j.jcp.2004.07.007}.

\bibitem[{Diamessis and Redekopp(2006)Diamessis, and Redekopp}]{diamessis2006}
Diamessis, P., and L.~Redekopp, 2006: Numerical investigation of solitary
  internal wave-induced global instability in a shallow water benthic boundary
  layers. \textit{J.\ Phys.\ Oceanogr.}, \textbf{36}, 784--812,
  \doi{10.1175/JPO2900.1}.

\bibitem[{Duda et~al.(2004)Duda, Lynch, Irish, Beardsley, Ramp, Chiu, Tang,,
  and Yang}]{duda2004}
Duda, T., J.~Lynch, J.~Irish, R.~Beardsley, S.~Ramp, C.-S. Chiu, T.-Y. Tang,
  and Y.-J. Yang, 2004: Internal tide and nonlinear internal wave behavior at
  the continental slope in the {N}orthern {S}outh {C}hina {S}ea. \textit{IEEE
  J.\ Oceanic Eng.}, \textbf{29~(4)}, \doi{10.1109/JOE.2004.836998}.

\bibitem[{Dunphy et~al.(2011)Dunphy, Subich,, and Stastna}]{dunphy2011}
Dunphy, M., C.~Subich, and M.~Stastna, 2011: Spectral methods for internal
  waves: indistinguishable density profiles and double-humped solitary waves.
  \textit{Nonlinear Proc.\ Geoph.}, \textbf{18}, 351--358,
  \doi{10.5194/npg-18-351-2011}.

\bibitem[{Fringer and Street(2003)Fringer, and Street}]{fringer2003}
Fringer, O., and R.~Street, 2003: The dynamics of breaking progressive
  interfacial waves. \textit{J.\ Fluid\ Mech.}, \textbf{494}, 319--353,
  \doi{10.1017/S0022112003006189}.

\bibitem[{Fructus et~al.(2009)Fructus, Carr, Grue, Jensen,, and
  Davies}]{fructus2009}
Fructus, D., M.~Carr, J.~Grue, A.~Jensen, and P.~Davies, 2009: Shear-induced
  breaking of large internal solitary waves. \textit{J.\ Fluid\ Mech.},
  \textbf{18}, 351--358, \doi{10.1017/S0022112008004898}.

\bibitem[{Gottlieb and Hesthaven(2001)Gottlieb, and Hesthaven}]{gottlieb2001}
Gottlieb, D., and J.~Hesthaven, 2001: Spectral methods for hyperbolic problems.
  \textit{J. Comp. Appl. Math.}, \textbf{128~(1)}, 83--131,
  \doi{10.1016/bs.hna.2016.09.007}, numerical Analysis 2000. Vol. VII: Partial
  Differential Equations.

\bibitem[{He et~al.(2019)He, Lamb,, and Lien}]{he2019}
He, Y., K.~Lamb, and R.-C. Lien, 2019: Internal solitary waves with subsurface
  cores. \textit{J.\ Fluid\ Mech.}, \textbf{873}, 1--17,
  \doi{10.1017/jfm.2019.407}.

\bibitem[{Helfrich(1992)}]{helfrich1992}
Helfrich, K., 1992: Internal solitary wave breaking and run-up on a uniform
  slope. \textit{J.\ Fluid\ Mech.}, \textbf{243}, 133--154,
  \doi{10.1017/S0022112092002660}.

\bibitem[{Helfrich and Melville(2006)Helfrich, and Melville}]{helfrich2006}
Helfrich, K., and W.~Melville, 2006: Long nonlinear internal waves.
  \textit{Annu.\ Rev.\ Fluid Mech.}, \textbf{38}, 395--425,
  \doi{10.1146/annurev.fluid.38.050304.092129}.

\bibitem[{Huang et~al.(2022)Huang, Huang, Zhao, Chang, Xu, Yang,, and
  Tian}]{Huang2022}
Huang, S., X.~Huang, W.~Zhao, Z.~Chang, X.~Xu, Q.~Yang, and J.~Tian, 2022:
  Shear instability in internal solitary waves in the {N}orthern {S}outh
  {C}hina {S}ea induced by multiscale background processes. \textit{J. Phys.
  Oceanogr.}, \textbf{52~(12)}, 2975 -- 2994, \doi{10.1175/JPO-D-21-0241.1}.

\bibitem[{Jackson et~al.(2012)Jackson, da~Silva,, and Jeans}]{jackson2012}
Jackson, C., J.~da~Silva, and G.~Jeans, 2012: The generation of nonlinear
  internal waves. \textit{Oceanography}, \textbf{25},
  \doi{10.5670/oceanog.2012.46}.

\bibitem[{Jacobitz et~al.(1997)Jacobitz, Sarkar,, and Van~Atta}]{jacobitz1997}
Jacobitz, F., S.~Sarkar, and C.~Van~Atta, 1997: Direct numerical simulations of
  the turbulence evolution in a uniformly sheared and stably stratified flow.
  \textit{J.\ Fluid\ Mech.}, \textbf{342}, 231--261.

\bibitem[{Joshi(2016)}]{joshi2016c}
Joshi, S., 2016: Development of fast high-order numerical methods for
  high-{R}eynolds number environmental flows. Ph.D. thesis, Cornell University.

\bibitem[{Joshi et~al.(2016{\natexlab{a}})Joshi, Diamessis, Steinmoller,
  Stastna,, and Thomsen}]{joshi2016b}
Joshi, S., P.~Diamessis, D.~Steinmoller, M.~Stastna, and G.~Thomsen,
  2016{\natexlab{a}}: A post-processing technique for stabilizing the
  discontinuous pressure projection operator in marginally-resolved
  incompressible inviscid flow. \textit{Computers and Fluids},
  \textbf{~(1-10)}.

\bibitem[{Joshi et~al.(2016{\natexlab{b}})Joshi, Thomsen,, and
  Diamessis}]{joshi2016a}
Joshi, S., G.~Thomsen, and P.~Diamessis, 2016{\natexlab{b}}:
  Deflation-accelerated preconditioning of the {P}oisson–{N}eumann {S}chur
  problem on long domains with a high-order discontinuous element-based
  collocation method. \textit{J.\ Comput.\ Phys.}, \textbf{313~(209-232)},
  \doi{https://doi.org/10.1016/j.jcp.2016.02.033}.

\bibitem[{Karniadakis et~al.(1991)Karniadakis, Israeli,, and
  Orszag}]{karniadakis1991}
Karniadakis, G., M.~Israeli, and S.~Orszag, 1991: High-order splitting methods
  for the incompressible navier-stokes equations. \textit{J.\ Comput.\ Phys.},
  \textbf{97}, 411--443, \doi{10.1016/0021-9991(91)90007-8}.

\bibitem[{Karniadakis and Sherwin(2005)Karniadakis, and
  Sherwin}]{karniadakis2005}
Karniadakis, G., and S.~Sherwin, 2005: \textit{Spectral/HP Element Methods for
  Computational Fluid Dynamics}.
  \doi{10.1093/acprof:oso/9780198528692.001.0001}.

\bibitem[{Kirby and Karniadakis(2003)Kirby, and Karniadakis}]{kirby2003}
Kirby, R., and G.~Karniadakis, 2003: De-aliasing on non-uniform grids:
  Algorithms and applications. \textit{J.\ Comput.\ Phys.}, \textbf{191},
  249--264, \doi{10.1016/S0021-9991(03)00314-0}.

\bibitem[{Klose et~al.(2020)Klose, Jacobs,, and Serra}]{klose2020}
Klose, B.~F., G.~B. Jacobs, and M.~Serra, 2020: Kinematics of lagrangian flow
  separation in external aerodynamics. \textit{AIAA Journal}, \textbf{58~(5)},
  1926--1938, \doi{10.2514/1.J059026}.

\bibitem[{Klymak and Moum(2003)Klymak, and Moum}]{klymak2003}
Klymak, J., and J.~Moum, 2003: Internal solitary waves of elevation advancing
  on a shoaling shelf. \textit{Geophys.\ Res.\ Lett.}, \textbf{30~(20)},
  \doi{10.1029/2003GL017706}, 2045.

\bibitem[{Koop and Butler(1981)Koop, and Butler}]{koop1981}
Koop, G., and G.~Butler, 1981: An investigation of internal solitary waves in a
  two-fluid system. \textit{J.\ Fluid\ Mech.}, \textbf{112}, 225--251,
  \doi{10.1017/S0022112081000372}.

\bibitem[{Kopriva(2009)}]{kopriva}
Kopriva, D., 2009: \textit{Implementing Spectral Methods for Partial
  Differential Equations}. Springer.

\bibitem[{Lamb(2002)}]{lamb2002}
Lamb, K., 2002: A numerical investigation of solitary internal waves with
  trapped cores formed via shoaling. \textit{J.\ Fluid\ Mech.}, \textbf{451},
  109--144, \doi{10.1017/S002211200100636X}.

\bibitem[{Lamb(2003)}]{lamb2003}
Lamb, K., 2003: Shoaling solitary internal waves: on a criterion for the
  formation of waves with trapped cores. \textit{J.\ Fluid\ Mech.},
  \textbf{478}, 81--100, \doi{10.1017/S0022112002003269}.

\bibitem[{Lamb and Farmer(2011)Lamb, and Farmer}]{lamb2011}
Lamb, K., and D.~Farmer, 2011: Instabilities in an internal solitary-like wave
  on the {O}regon {S}helf. \textit{J.\ Phys.\ Oceanogr.}, \textbf{41}, 67--87,
  \doi{https://doi.org/10.1175/2010JPO4308.1}.

\bibitem[{Lamb and Warn-Varnas(2015)Lamb, and Warn-Varnas}]{lamb2015}
Lamb, K., and A.~Warn-Varnas, 2015: Two-dimensional numerical simulations of
  shoaling internal solitary waves at the {ASIAEX} site in the {S}outh {C}hina
  {S}ea. \textit{Nonlinear Proc.\ Geoph.}, \textbf{22}, 289--312,
  \doi{https://doi.org/10.5194/npg-22-289-2015}.

\bibitem[{Lamb et~al.(2019)Lamb, Lien,, and Diamessis}]{lamb2019}
Lamb, K.~G., R.-C. Lien, and P.~J. Diamessis, 2019: Internal solitary waves and
  mixing. \textit{Encyclopedia of Ocean Sciences (Third Edition)}, J.~K.
  Cochran, H.~J. Bokuniewicz, and P.~L. Yager, Eds., third edition ed.,
  Academic Press, Oxford, 533--541,
  \doi{https://doi.org/10.1016/B978-0-12-409548-9.10951-0}.

\bibitem[{Legg et~al.(2009)}]{legg:09}
Legg, S., and Coauthors, 2009: Improving oceanic overflow representation in
  climate models: The gravity current entrainment climate process team.
  \textit{Bulletin Am. Met. Soc.}, \textbf{90~(5)}, 657--670,
  \doi{10.1175/2008BAMS2667.1}.

\bibitem[{Lien et~al.(2012)Lien, D'Asaro, Henyey, Chang, Tang,, and
  Yang}]{lien2012}
Lien, R.-C., E.~D'Asaro, F.~Henyey, M.-H. Chang, T.-Y. Tang, and Y.-J. Yang,
  2012: Trapped core formation within a shoaling nonlinear internal wave.
  \textit{J.\ Phys.\ Oceanogr.}, \textbf{42}, 511--525,
  \doi{10.1175/2011JPO4578.1}.

\bibitem[{Lien et~al.(2014)Lien, Henyey, Ma,, and Yang}]{lien2014}
Lien, R.-C., F.~Henyey, B.~Ma, and Y.-J. Yang, 2014: Large-amplitude internal
  solitary waves observed in the northern {S}outh {C}hina {S}ea: Properties and
  energetics. \textit{J.\ Phys.\ Oceanogr.}, \textbf{44~(4)}, 1095--1115,
  \doi{10.1175/JPO-D-13-088.1}.

\bibitem[{Lien et~al.(2005)Lien, Yang, Chang,, and D'Asaro}]{lien2005}
Lien, R.-C., T.~Yang, M.~Chang, and E.~D'Asaro, 2005: Energy of nonlinear
  internal waves in the {S}outh {C}hina {S}ea. \textit{Geophys.\ Res.\ Lett.},
  \textbf{32~(L05615)}, \doi{10.1029/2004GL022012}.

\bibitem[{Linden(2012)}]{linden2012}
Linden, P., 2012: \textit{Gravity currents – theory and laboratory
  experiments}, 13–51. Cambridge University Press.

\bibitem[{Lloret et~al.(2024)Lloret, Diamessis, Stastna,, and
  Thomsen}]{lloret:24b}
Lloret, P., P.~J. Diamessis, M.~Stastna, and G.~N. Thomsen, 2024: A robust
  numerical method for the generation and propagation of periodic
  finite-amplitude internal waves in natural waters using high-accuracy
  simulations. \textit{Nonlin. Proc. Geoph.}, \textbf{31~(4)}, 515--533,
  \doi{10.5194/npg-31-515-2024}.

\bibitem[{Long(1953)}]{long1953}
Long, R., 1953: Some aspects of the flow of stratified fluids i. a theoretical
  investigation. \textit{Tellus}, \textbf{8}, 460--471,
  \doi{10.1111/j.2153-3490.1955.tb01171.x}.

\bibitem[{Lucas and Pinkel(2022)Lucas, and Pinkel}]{lucas2022}
Lucas, A.~J., and R.~Pinkel, 2022: Observations of coherent transverse wakes in
  shoaling nonlinear internal waves. \textit{J. Phys. Oceanogr.},
  \textbf{52~(6)}, \doi{10.1175/JPO-D-21-0059.1}.

\bibitem[{Malm et~al.(2013)Malm, Schlatter, Fischer,, and
  Henningson}]{malm2013}
Malm, J., P.~Schlatter, P.~Fischer, and D.~Henningson, 2013: Stabilization of
  the spectral element method in convection dominated flows by recovery of
  skew-symmetry. \textit{J. Sci. Comput.}, \textbf{57},
  \doi{10.1007/s10915-013-9704-1}.

\bibitem[{Marques et~al.(2017)Marques, Wells, Padman,, and
  Ozgokmen}]{marques:17}
Marques, G.~M., M.~G. Wells, L.~Padman, and T.~M. Ozgokmen, 2017: Flow
  splitting in numerical simulations of oceanic dense-water outflows.
  \textit{Ocean Modeling}, \textbf{113}, 66--84,
  \doi{10.1016/j.ocemod.2017.03.011}.

\bibitem[{Mashayek and Peltier(2011)Mashayek, and Peltier}]{mashayek:11}
Mashayek, A., and W.~R. Peltier, 2011: Three-dimensionalization of the
  stratified mixing layer at high reynolds number. \textit{Phys. Fluids},
  \textbf{23~(11)}, \doi{10.1063/1.3651269}.

\bibitem[{Moore and Lien(2007)Moore, and Lien}]{moore2007}
Moore, S., and R.-C. Lien, 2007: Pilot whales follow internal solitary waves in
  the {S}outh {C}hina {S}ea. \textit{Mar.\ Mammal Sci.}, \textbf{21~(1)},
  193--196, \doi{10.1111/j.1748-7692.2006.00086.x}.

\bibitem[{Moum et~al.(2003)Moum, Farmer, Smyth, Armi,, and Vagle}]{moum2003}
Moum, J., D.~Farmer, W.~Smyth, L.~Armi, and S.~Vagle, 2003: Structure and
  generation of turbulence at interfaces strained by internal solitary waves
  propagating shoreward over the continental shelf. \textit{J.\ Phys.\
  Oceanogr.}, \textbf{33}, 2093--2112,
  \doi{10.1175/1520-0485(2003)033<2093:SAGOTA>2.0.CO;2}.

\bibitem[{Moum et~al.(2007)Moum, Farmer, Shroyer, Smyth,, and Armi}]{moum2007a}
Moum, J.~N., D.~M. Farmer, E.~L. Shroyer, W.~D. Smyth, and L.~Armi, 2007:
  Dissipative losses in nonlinear internal waves propagating across the
  continental shelf. \textit{J.\ Phys.\ Oceanogr.}, \textbf{37~(7)},
  1989--1995, \doi{10.1175/JPO3091.1}.

\bibitem[{Olsthoorn et~al.(2023)Olsthoorn, Kaminski,, and Robb}]{Olsthoorn2023}
Olsthoorn, J., A.~K. Kaminski, and D.~M. Robb, 2023: Dynamics of asymmetric
  stratified shear instabilities. \textit{Phys. Rev. Fluids}, \textbf{8},
  024\,501, \doi{10.1103/PhysRevFluids.8.024501}.

\bibitem[{Ozgokmen et~al.(2009)Ozgokmen, Iliescu,, and Fischer}]{ozgokmen:09}
Ozgokmen, T.~M., T.~Iliescu, and P.~F. Fischer, 2009: Large eddy simulation of
  stratified mixing in a three-dimensional lock-exchange system. \textit{Ocean
  Modeling}, \textbf{26}, 134--1556.

\bibitem[{Passaggia et~al.(2018)Passaggia, Helfrich,, and
  White}]{Passaggia_2018}
Passaggia, P.-Y., K.~R. Helfrich, and B.~L. White, 2018: Optimal transient
  growth in thin-interface internal solitary waves. \textit{J.\ Fluid\ Mech.},
  \textbf{840}, 342–378, \doi{10.1017/jfm.2018.19}.

\bibitem[{Pope(2000)}]{pope}
Pope, S., 2000: \textit{Turbulent Flows}. Cambridge University Press,
  \doi{10.1016/S0010-2180(01)00244-9}.

\bibitem[{Ramp et~al.(2022{\natexlab{a}})Ramp, Yang, Chiu, Reeder,, and
  Bahr}]{Ramp2022a}
Ramp, S., Y.~Yang, C.-S. Chiu, D.~Reeder, and F.~Bahr, 2022{\natexlab{a}}:
  Observations of shoaling internal wave transformation over a gentle slope in
  the {S}outh {C}hina {S}ea. \textit{Nonlinear Proc.\ Geoph.}, \textbf{29},
  279--299, \doi{10.5194/npg-29-279-2022}.

\bibitem[{Ramp et~al.(2022{\natexlab{b}})}]{Ramp2022b}
Ramp, S.~R., and Coauthors, 2022{\natexlab{b}}: Solitary waves impinging on an
  isolated tropical reef: Arrival patterns and wave transformation under
  shoaling. \textit{J. Geophys. Res.: Oceans}, \textbf{127~(3)},
  \doi{10.1029/2021JC017781}.

\bibitem[{Rivera-Rosario et~al.(2022)Rivera-Rosario, Diamessis, Lien, Lamb,,
  and Thomsen}]{riverarosario2022}
Rivera-Rosario, G., P.~Diamessis, R.-C. Lien, K.~Lamb, and G.~Thomsen, 2022:
  Three-dimensional perspective on a convective instability and transition to
  turbulence in an internal solitary wave of depression shoaling over gentle
  slopes. \textit{Env. Fluid Mech.}, \doi{10.1007/s10652-022-09844-7}.

\bibitem[{Rivera-Rosario et~al.(2020)Rivera-Rosario, Diamessis, Lien, Lamb,,
  and Thomsen}]{riverarosario2020}
Rivera-Rosario, G., P.~J. Diamessis, R.-C. Lien, K.~G. Lamb, and G.~N. Thomsen,
  2020: {Formation of Recirculating Cores in Convectively Breaking Internal
  Solitary Waves of Depression Shoaling over Gentle Slopes in the South China
  Sea}. \textit{J. Phys. Oceanogr.}, \textbf{50~(5)}, 1137--1157,
  \doi{10.1175/JPO-D-19-0036.1}.

\bibitem[{Sakai et~al.(2020)Sakai, Diamessis,, and Redekopp}]{sakai:20a}
Sakai, T., P.~J. Diamessis, and L.~G. Redekopp, 2020: Self-sustained
  instability, transition and turbulence induced by a long separation bubble in
  the footprint of an internal solitary wave { Part I: Flow topology}.
  \textit{Phys. Rev. Fluids}, \textbf{5}, 103\,801.

\bibitem[{Salehipour et~al.(2016)Salehipour, Caulfield,, and
  Peltier}]{salehipour2016}
Salehipour, H., C.~P. Caulfield, and W.~R. Peltier, 2016: Turbulent mixing due
  to the {H}olmboe wave instability at high {R}eynolds number. \textit{J.\
  Fluid\ Mech.}, \textbf{803}, 591–621, \doi{10.1017/jfm.2016.488}.

\bibitem[{Salehipour et~al.(2018)Salehipour, Peltier,, and
  Caulfield}]{salehipour2018}
Salehipour, H., W.~Peltier, and C.~Caulfield, 2018: Self-organized criticality
  of turbulence in strongly stratified mixing layers. \doi{10.17863/CAM.33527}.

\bibitem[{Sandstrom and Elliott(1984)Sandstrom, and Elliott}]{sandstrom1984}
Sandstrom, H., and J.~Elliott, 1984: Internal tide and solitons on the
  {S}cotian shelf: A nutrient pump at work. \textit{J.\ Geophys.\ Res.},
  \textbf{89~(C4)}, 6415--6426, \doi{10.1029/JC089iC04p06415}.

\bibitem[{Schmid and Henningson(2001)Schmid, and Henningson}]{schmid2001}
Schmid, P., and D.~Henningson, 2001: \textit{Stability and Transition in Shear
  Flows}, Vol. 142. \doi{10.1007/978-1-4613-0185-1}.

\bibitem[{Scotti and Mitran(2008)Scotti, and Mitran}]{Scotti:08}
Scotti, A., and S.~Mitran, 2008: {An approximated method for the solution of
  elliptic problems in thin domains: Application to nonlinear internal waves}.
  \textit{Ocean Modelling}, \textbf{25}, 144--153,
  \doi{10.1016/j.ocemod.2008.07.005}.

\bibitem[{Scotti and Pineda(2004)Scotti, and Pineda}]{scotti2004}
Scotti, A., and J.~Pineda, 2004: Observation of very large and steep internal
  waves of elevation near the massachusetts coast. \textit{Geophys.\ Res.\
  Lett.}, \textbf{31~(22)}, \doi{10.1029/2004GL021052}, l22307.

\bibitem[{Shroyer et~al.(2011)Shroyer, Moum,, and Nash}]{shroyer2011}
Shroyer, E., J.~Moum, and J.~Nash, 2011: Nonlinear internal waves over {N}ew
  {J}ersey's continental shelf. \textit{J.\ Geophys.\ Res.},
  \textbf{116~(C03022)}, \doi{10.1029/2010JC006332}.

\bibitem[{Sinnett et~al.(2022)Sinnett, Ramp, Yang, Chang, Jan,, and
  Davis}]{Sinnett2022}
Sinnett, G., S.~R. Ramp, Y.~J. Yang, M.-H. Chang, S.~Jan, and K.~A. Davis,
  2022: Large-amplitude internal wave transformation into shallow water.
  \textit{J. Phys. Oceanogr.}, \textbf{52~(10)}, 2539 -- 2554,
  \doi{10.1175/JPO-D-21-0273.1}.

\bibitem[{Smyth and Moum(2000)Smyth, and Moum}]{smyth2000}
Smyth, W., and J.~Moum, 2000: Length scales of turbulence in stably stratified
  mixing layers. \textit{Phys.\ Fluids}, \textbf{12~(6)}, 1327--1342.

\bibitem[{Smyth(2004)}]{Smyth2004}
Smyth, W.~D., 2004: Kelvin–helmholtz billow evolution from a localized
  source. \textit{Q. J. Roy. Meteorol. Soc.}, \textbf{130~(603)}, 2753--2766,
  \doi{https://doi.org/10.1256/qj.03.226}.

\bibitem[{Smyth and Carpenter(2019)Smyth, and Carpenter}]{Smyth_Carpenter_2019}
Smyth, W.~D., and J.~R. Carpenter, 2019: \textit{Instability in Geophysical
  Flows}. Cambridge University Press, \doi{10.1017/9781108640084}.

\bibitem[{Stastna(2022)}]{Stastna2022}
Stastna, M., 2022: \textit{Internal Waves in the Ocean: Theory and Practice}.
  Springer Cham, \doi{10.1007/978-3-030-99210-1}.

\bibitem[{Stastna and Lamb(2002)Stastna, and Lamb}]{stastna2002}
Stastna, M., and K.~Lamb, 2002: Large fully nonlinear internal solitary waves:
  The effect of background current. \textit{Phys.\ Fluids}, \textbf{14~(9)},
  2987--2999, \doi{10.1063/1.1496510}.

\bibitem[{Stastna and Lamb(2008)Stastna, and Lamb}]{stastna2008}
Stastna, M., and K.~Lamb, 2008: Sediment resuspension mechanism associated with
  internal waves in coastal waters. \textit{J.\ Geophys.\ Res.}, \textbf{113},
  \doi{10.1146/annurev-fluid-122316-045049}.

\bibitem[{Stastna and Legare(2024)Stastna, and Legare}]{stastna2024}
Stastna, M., and S.~Legare, 2024: Simulations of shoaling large-amplitude
  internal waves: perspectives and outlook. \textit{Flow}, \textbf{4},
  \doi{10.1017/flo.2024.9}.

\bibitem[{Tennekes and Lumley(1972)Tennekes, and Lumley}]{tennekes}
Tennekes, H., and J.~Lumley, 1972: \textit{A First Course in Turbulence}. MIT
  Press, \doi{10.1017/S002211207321251X}.

\bibitem[{Thyng et~al.(2016)Thyng, Greene, Hetland, Zimmerle,, and
  Dimarco}]{thyng2016}
Thyng, K., C.~Greene, R.~Hetland, H.~Zimmerle, and S.~Dimarco, 2016: True
  colors of oceanography: Guidelines for effective and accurate colormap
  selection. \textit{Oceanography}, \textbf{29}, 9--13,
  \doi{10.5670/oceanog.2016.66}.

\bibitem[{Towns et~al.(2014)}]{XSEDE}
Towns, J., and Coauthors, 2014: Xsede: Accelerating scientific discovery.
  \textit{Computing in Science \& Engineering}, \textbf{16~(5)}, 62--74,
  \doi{10.1109/MCSE.2014.80}.

\bibitem[{Troy and Koseff(2005)Troy, and Koseff}]{troy2005}
Troy, C., and J.~Koseff, 2005: The instability and breaking of long internal
  waves. \textit{J.\ Fluid\ Mech.}, \textbf{543}, 107--136,
  \doi{10.1017/S0022112005006798}.

\bibitem[{Turkington et~al.(1991)Turkington, Eydeland,, and
  Wang}]{turkington1991}
Turkington, B., A.~Eydeland, and S.~Wang, 1991: A computational method for
  solitary internal waves in a continously stratified fluid. \textit{Stud.
  Appl.\ Math.}, \textbf{85}, 93--127, \doi{10.1002/sapm199185293}.

\bibitem[{Vladoiu et~al.(2022)Vladoiu, Lien,, and Kunze}]{vladoiu:22}
Vladoiu, A., R.-C. Lien, and E.~Kunze, 2022: Two-dimensional wavenumber spectra
  on the horizontal submesoscale and vertical finescale. \textit{J. Phys.
  Oceanogr.}, \textbf{52~(9)}, 2009--2028, \doi{10.1175/JPO-D-21-0111.1}.

\bibitem[{Vlasenko et~al.(2005)Vlasenko, Ostrovsky,, and Hutter}]{vlasenko2005}
Vlasenko, V., L.~Ostrovsky, and K.~Hutter, 2005: Adiabatic behavior of strongly
  nonlinear internal solitary waves in slope-shelf areas. \textit{J.\ Geophys.\
  Res.}, \textbf{110~(C04006)}, \doi{10.1029/2004JC002705}.

\bibitem[{Vlasenko et~al.(2006)Vlasenko, Stashchuck,, and
  Hutter}]{vlasenko2005b}
Vlasenko, V., N.~Stashchuck, and K.~Hutter, 2006: Review of baroclinic tides:
  Theoretical modeling and observational evidence. \textit{Baroclinic Tides,
  pp. 372. ISBN 0521843952. Cambridge, UK: Cambridge University Press, July
  2005.}, \doi{10.1017/CBO9780511535932}.

\bibitem[{Whitwell et~al.(2024)Whitwell, Jones, Ivey, Rosevear,, and
  Rayson}]{Whitwell2024}
Whitwell, C.~A., N.~L. Jones, G.~N. Ivey, M.~G. Rosevear, and M.~D. Rayson,
  2024: Ocean mixing in a shelf sea driven by energetic internal waves.
  \textit{J. Geophys. Res.: Oceans}, \textbf{129~(2)}, e2023JC019\,704,
  \doi{10.1029/2023JC019704}, e2023JC019704 2023JC019704.

\bibitem[{Winters and D'Asaro(1994)Winters, and D'Asaro}]{winters1994}
Winters, K., and E.~D'Asaro, 1994: Three-dimensional wave instability near a
  critical layer. \textit{J.\ Fluid\ Mech.}, \textbf{272}, 255--284,
  \doi{10.1017/S0022112094004465}.

\bibitem[{Winters et~al.(1995)Winters, Lombard, Riley,, and
  D'Asaro}]{winters1995}
Winters, K., P.~Lombard, J.~Riley, and E.~D'Asaro, 1995: Available potential
  energy and mixing in density-stratified fluids. \textit{J.\ Fluid\ Mech.},
  \textbf{289}, 115--128, \doi{10.1017/S002211209500125X}.

\bibitem[{Winters(2015)}]{winters:15}
Winters, K.~B., 2015: Tidally driven mixing and dissipation in the stratified
  boundary layer above steep submarine topography. \textit{Geophys. Res.
  Lett.}, \textbf{42~(17)}, 7123--7130,
  \doi{https://doi.org/10.1002/2015GL064676}.

\bibitem[{Winters et~al.(2004)Winters, McKinnon,, and Mills}]{winters:04}
Winters, K.~B., J.~McKinnon, and B.~Mills, 2004: A spectral model for process
  studies of density stratified flows. \textit{J. of Atmos. Ocean. Techn.},
  \textbf{21~(1)}, 69--94.

\bibitem[{Winters and Riley(1992)Winters, and Riley}]{winters1992}
Winters, K.~B., and J.~J. Riley, 1992: Instability of internal waves near a
  critical level. \textit{Dyn.\ Atmos.\ Oceans}, \textbf{16}, 249--278,
  \doi{10.1016/0377-0265(92)90009-I}.

\bibitem[{Xu et~al.(2019)Xu, Stastna,, and Deepwell}]{Xu2019}
Xu, C., M.~Stastna, and D.~Deepwell, 2019: Spontaneous instability in internal
  solitary-like waves. \textit{Phys. Rev. Fluids}, \textbf{4}, 014\,805,
  \doi{10.1103/PhysRevFluids.4.014805}.

\bibitem[{Yeh et~al.(2024)Yeh, Chang, Lien, Chang, Chen, Jan, Yang,, and
  Vladoiu}]{yeh:24}
Yeh, Y., M.~Chang, R.~Lien, J.~Chang, J.~Chen, S.~Jan, Y.~J. Yang, and
  A.~Vladoiu, 2024: Turbulence generation via nonlinear lee wave trailing edge
  instabilities in kuroshio-seamount interactions. \textit{J. Geophys. Res.},
  \textbf{129~(9)}, \doi{10.1029/2024JC020971}.

\bibitem[{Zhang and Alford(2015)Zhang, and Alford}]{zhang2015}
Zhang, S., and M.~Alford, 2015: Instabilities in nonlinear internal waves on
  the {W}ashington continental shelf. \textit{J.\ Geophys.\ Res.},
  \textbf{120}, 5272--5283, \doi{10.1002/2014JC010638}.

\bibitem[{Zhang and Samtaney(2016)Zhang, and Samtaney}]{zhang:16}
Zhang, W., and R.~Samtaney, 2016: Assessment of spanwise domain size effect on
  the transitional flow past an airfoil. \textit{Comp. \& Fluids},
  \textbf{124}, 39--53, \doi{10.1016/j.compfluid.2015.10.008}.

\end{thebibliography}

\end{document}